%% file: vcc.tex
\def\conf{0}
\def\hideappendix{0}
\InputIfFileExists{SODA_conf}{}{}
\documentclass[11pt,letterpaper]{article}
\usepackage{times}
\usepackage{fullpage,amssymb,amsmath}
\usepackage[ruled,vlined,linesnumbered]{algorithm2e}
\usepackage{epsfig,color}

\input{pream}

\SetKw{KwDownTo}{downto}

\dontprintsemicolon

\begin{document}

\begin{titlepage}
\title{A Near-Optimal Sublinear-Time Algorithm\\
 for Approximating the Minimum Vertex Cover Size\\
\ifnum\conf=1
\Large\textit{(Extended Abstract)}
\fi
}

\author{Krzysztof Onak\thanks{School of Computer Science, Carnegie Mellon University,
E-mail: {\tt konak@cs.cmu.edu}. Research supported by a Simons Postdoctoral Fellowship and
the NSF grant 0728645.}
\and Dana Ron \thanks {School of Electrical Engineering, Tel Aviv
University.
E-mail: {\tt danar@eng.tau.ac.il}.
Research supported by the Israel Science Foundation grant number 246/08.}
\and Michal Rosen \thanks{Blavatnik School of Computer Science, Tel Aviv University,
 E-mail: {\tt  micalros@post.tau.ac.il}}
\and
Ronitt Rubinfeld\thanks{CSAIL, MIT, Cambridge MA 02139 and
the Blavatnik School of Computer Science, Tel Aviv University.
 E-mail: {\tt  ronitt@csail.mit.edu}.
Research supported by NSF awards CCF-1065125 and CCF-0728645, Marie Curie Reintegration grant PIRG03-GA-2008-231077 and the Israel Science Foundation grant nos.\ 1147/09 and  1675/09.}
}

\maketitle

\begin{abstract}
We give a nearly optimal sublinear-time algorithm for
approximating the size of a minimum vertex cover in a graph $G$.
The algorithm may query the degree $\deg(v)$ of any vertex $v$ of its choice,
and for each $1 \leq i \leq \deg(v)$, it may ask for the $i^{\rm th}$ neighbor
of $v$. Letting $\VCopt(G)$ denote the minimum size of vertex cover in $G$, the algorithm
outputs, with high constant success probability, an estimate  $\wVC(G)$
such that $\VCopt(G) \leq \wVC(G) \leq 2 \VCopt(G) + \eps n$, where $\eps$ is
a given additive approximation parameter. We refer to such an estimate as
a $(2,\eps)$-estimate. The query complexity and running
time of the algorithm are $\tilde{O}(\bd \cdot {\rm poly}(1/\eps))$, where
$\bd$ denotes the average vertex degree in the graph.
The best previously known sublinear algorithm, of Yoshida et al.\ ({\em STOC 2009\/}),
has query complexity and running time $O(d^4/\eps^2)$, where $d$ is the maximum degree in the graph.
Given the lower bound of $\Omega(\bd)$ (for constant $\eps$) for obtaining such an estimate
(with any constant multiplicative factor) due to Parnas and Ron ({\em TCS 2007\/}),
our result is nearly optimal.

In the case that the graph is dense, that is, the number of edges is $\Theta(n^2)$,
we consider another model,
 in which the algorithm may ask, for any pair of vertices $u$ and $v$, whether
there is an edge between $u$ and~$v$. We show how to adapt the algorithm
that uses neighbor queries to this model
and obtain an
algorithm that outputs a $(2,\eps)$-estimate of the size of
a minimum vertex cover whose query complexity and running
time are $\tilde{O}(n) \cdot {\rm \poly}(1/\eps)$.

\end{abstract}

\thispagestyle{empty}
\end{titlepage}

\input{intro.tex}

\input{alg}

\input{mm-short}

\ifnum\conf=0
\input{mm}

\fi

\input{near-opt}

\bibliographystyle{alpha}
\bibliography{vc}

\ifnum\conf=1
\ifnum\hideappendix=1
\end{document}
\fi
\fi

\ifnum\conf=1
\section{Missing Details for Section~\ref{avg-calls.sec}}\label{avg-calls.app}

\input{alg2.tex}
\section{Missing Details for Section~\ref{mm.sec}}\label{mm.app}

\input{mm.tex}
\section{Missing Details for Section~\ref{near-opt.sec}}\label{near-opt.app}

\NearOpt
\Dense

\fi
\end{document}

%% file: pream.tex
\ifnum\conf=0
\fi

\newtheorem{theorem}{Theorem}[section]

\newtheorem{definition}[theorem]{Definition}

\newtheorem{claim}[theorem]{Claim}
\newtheorem{lemma}[theorem]{Lemma}

\newtheorem{corollary}[theorem]{Corollary}

\newtheorem{define}[theorem]{Definition}

\newtheorem{obser}[theorem]{Observation}

\newcommand{\BE}{\begin{enumerate}} \newcommand{\EE}{\end{enumerate}}
\newcommand{\BI}{\begin{itemize}} \newcommand{\EI}{\end{itemize}}
\newcommand{\BDes}{\begin{description}}\newcommand{\EDes}{\end{description}}
\newcommand{\BT}{\begin{theorem}} \newcommand{\ET}{\end{theorem}}
\newcommand{\BL}{\begin{lemma}} \newcommand{\EL}{\end{lemma}}
\newcommand{\BD}{\begin{define}} \newcommand{\ED}{\end{define}}
\newcommand{\BCM}{\begin{claim}} \newcommand{\ECM}{\end{claim}}
\newcommand{\BC}{\begin{corollary}} \newcommand{\EC}{\end{corollary}}
\newcommand{\BA}{\begin{algorithm}} \newcommand{\EA}{\end{algorithm}}

\def\FullBox{\hbox{\vrule width 8pt height 8pt depth 0pt}}
\newcommand{\qed}{\;\;\;\FullBox}
\newenvironment{prf}{\noindent{\bf Proof:~~}}{\(\qed\)}
\newcommand{\BPF}{\begin{prf}} \newcommand {\EPF}{\end{prf}}
\newenvironment{proofof}[1]{\noindent{\bf Proof of {#1}:~}}{\endprf}
\newcommand{\BPFOF}{\begin{proofof}} \newcommand {\EPFOF}{\end{proofof}}

\newcommand{\BEQN}{\begin{eqnarray}}\newcommand{\EEQN}{\end{eqnarray}}
\newcommand{\BEQ}{\begin{equation}} \newcommand{\EEQ}{\end{equation}}

\newcommand{\poly}{{\rm poly}}

\renewcommand{\Pr}{{\rm Pr}}

\newcommand{\eqdef}{\stackrel{\rm def}{=}}

\newcommand{\eps}{\epsilon}

\renewcommand{\deg}{{\rm deg}}

\newcommand{\VC}{{\rm VC}}
\newcommand{\MM}{{\rm M}}
\newcommand{\C}{{\rm C}}
\newcommand{\wVC}{{\widehat{\rm VC}}}

\newcommand{\bd}{{\bar{d}}}

\newcommand{\tG}{{\widetilde{G}}}
\newcommand{\tV}{{\widetilde{V}}}
\newcommand{\tE}{{\widetilde{E}}}
\newcommand{\bG}{{\overline{G}}}
\newcommand{\bV}{{\overline{V}}}
\newcommand{\bE}{{\overline{E}}}
\newcommand{\bM}{{\overline{M}}}

\newcommand{\VO}{{\rm VO}}
\newcommand{\MO}{{\rm MO}}

\ifnum\conf=0
\newcommand{\mnote}[1]{ \marginpar{\tiny\bf
             \begin{minipage}[t]{0.5in}
               \raggedright #1
            \end{minipage}}}
\else
\newcommand{\mnote}[1]{}
\fi

\newcommand{\remove}[1]{}
\newcommand{\ignore}[1]{}

\newcommand{\ds}{\mbox{\tt neighbors}}
\newcommand{\lowest}{\mbox{\tt lowest}}
\newcommand{\setvalue}{\mbox{\tt set\_value}}
\newcommand{\getlb}{\mbox{\tt lower\_bound}}
\DeclareMathOperator{\length}{length}

\newcommand{\VCopt}{\VC_{\rm opt}}

%% file: intro.tex
\section{Introduction}
\label{intro.sec}

Computing the size of a minimum vertex cover in a graph is a classic
NP-hard problem.  However, one can approximate
the optimal value of the solution to within a multiplicative factor
of two, via a neat and simple algorithm whose running time
is linear in the size of the graph (this algorithm was
independently discovered by 
Gavril and Yanakakis, see e.g.~\cite{PS}). 

A natural question is
whether it is possible to obtain a good 
approximation for the \emph{size} of the optimal
vertex cover in time that is {\em sublinear\/}
in the size of the graph $G$.
Since achieving a pure multiplicative approximation is
easily seen to require linear time, we focus on algorithms that
compute an estimate $\wVC(G)$ such that with high constant probability, 
$\VCopt(G) \leq \wVC(G) \leq \alpha \cdot \VCopt(G) + \eps n$, for $\alpha \geq 1$ and
$0 \leq \eps < 1$, where 
$\VCopt(G)$ denotes the minimum size of a vertex cover in $G$.
We refer to such an estimate $\wVC(G)$ as an {\em $(\alpha,\eps)$-estimate\/}
of $\VCopt(G)$.  Observe that in the special case
when the vertex cover is very large, 
namely $\VCopt(G) = \Theta(n)$ (which happens for example when the maximum degree
and the average degree are of the same order), then 
an $(\alpha,\eps)$-estimate yields an
$(\alpha+O(\eps))$-multiplicative approximation.

Since an algorithm with complexity sublinear in the
size of the graph cannot even read the entire graph, 
it must have {\em query access\/} to the graph.    
In this work we consider two standard models of queries. In the first model,
the algorithm may query the degree $\deg(v)$ of any vertex 
$v$ of its choice, and it may also query the $i^{\rm th}$ neighbor of
$v$ (where the the order on the neighbors is arbitrary).
In the second model, 
more appropriate when the graph is stored
as an adjacency matrix,
the algorithm can check in a single query whether
there is an edge between two vertices $v$ and $w$ chosen by the algorithm.
We focus on the first model, but we eventually show that our 
algorithm can be modified to work in the second model as well.

\ifnum\conf=1
\vspace{-1.5ex}
\fi
\paragraph{Previous work.}
The aforementioned question was first posed by Parnas and Ron~\cite{PR}, who showed how to obtain
a $(2,\eps)$-estimate (for any given additive approximation parameter $\eps$)
in time $d^{O(\log d/\eps^3)}$, where $d$ is the maximum degree in the 
graph. 
The dependence on the maximum degree $d$ can actually be replaced
by a dependence on $\bd/\eps$, where $\bd$ is the average degree in the graph~\cite{PR}.
The upper bound of $d^{O(\log d/\eps^3)}$ was significantly improved in 
a sequence of papers~\cite{MR,NO,YYI}, where the best result due to
 Yoshida, Yamamoto, and Ito~\cite{YYI} (who analyze an algorithm proposed by
Nguyen and Onak~\cite{NO})
is an upper bound of $O(d^4/\eps^2)$.
Their analysis can also easily be adapted to give an upper bound of $O(\bd^4/\eps^4)$
for graphs with bounded average vertex degree $\bd$.

On the negative side, it was also proved in~\cite{PR} that at least
a linear dependence on the average degree, $\bd$, is necessary. Namely, $\Omega(\bd)$
queries are necessary for obtaining an $(\alpha,\eps)$-estimate
for any $\alpha \geq 1$ and $\eps < 1/4$, provided that $\bd = O(n/\alpha)$,
and in particular this is true for $\alpha = 2$.
We also mention that obtaining a $(2-\gamma,\eps)$-estimate for
any constant $\gamma$  requires a number of queries that grows at least
as the square root of the number of vertices~\cite[due to Trevisan]{PR}.

\ifnum\conf=1
\vspace{-1.5ex}
\fi
\paragraph{Our Result.} 
\sloppy
In this work we describe and analyze an algorithm
that computes a $(2,\eps)$-estimate of $\VCopt(G)$ in 
time $\tilde{O}(\bd)\cdot \poly(1/\eps)$.
Note that since the graph contains $\bd n/2$ edges, our running time is sublinear
for all values of $\bd$.
In particular, for graphs of constant average degree, the running time is
independent of the number of nodes and edges in the graph, whereas for general graphs
it is bounded by at most the square root of the number
of edges.  In view of the aforementioned
lower bound of $\Omega(\bd)$, our algorithm is optimal in terms of the
dependence on the average degree up to a polylogarithmic factor.
Since our algorithm builds on previous work, and in particular on the
algorithm proposed and analyzed in~\cite{NO,YYI}, we describe the latter algorithm first.\footnote{Yoshida
et al.~\cite{YYI} actually analyze an algorithm for approximating the size of
 a maximal independent set.
They then apply it to the line graph of a given graph $G$, so as to obtain an estimate
of the size of a maximal matching, and hence of a minimum vertex cover (with a
 multiplicative cost of $2$ in the quality of the estimate). For the sake of
simplicity, we describe their algorithm directly for a maximal matching (minimum vertex
cover).}
We refer to this algorithm as {\bf Approx-VC-I}.

\ifnum\conf=1
\vspace{-1.5ex}
\fi
\paragraph{The Algorithm  Approx-VC-I.}
Recall that the size of a minimum vertex cover is lower-bounded by the
size of any (maximal) matching in the graph, and is upper-bounded by twice the size of any
maximal matching. This is indeed the basis of the aforementioned factor-two
approximation algorithm, which runs in linear-time.
To estimate the size of an arbitrary such maximal matching, the algorithm 
follows the sampling paradigm of Parnas and Ron~\cite{PR}.
That is, the algorithm {\bf  Approx-VC-I} selects, uniformly, 
independently and at random, 
$\Theta(d^2/\eps^2)$ edges. 
For each edge selected, it calls a {\em maximal
matching oracle\/}, which we describe momentarily, where the oracle's answers
indicate whether or not the edge is in the maximal matching $\cal M$, for some
arbitrary maximal matching $\cal M$ (that is not a function 
of the queries to the oracle).
The algorithm then
outputs an estimate of the size of the maximal matching $\cal M$ 
(and hence of a minimum vertex cover)
based on the fraction of sampled edges for which the maximal matching oracle
returned a positive answer. The number of sampled edges ensures that
with high constant probability, the additive error of the estimate is
$O((\eps/d) m) \leq   \eps n$, where $m$ is the number of edges in the graph.

The main idea of the algorithm follows the idea suggested
in~\cite{NO} which is to
simulate the answers of the standard sequential greedy algorithm.
The greedy algorithm supposes a 
fixed ranking (ordering) of the edges in $G$, which uniquely
determines a maximal matching as follows: 
proceeding along the edges according to the order determined by
the ranking, add to the matching each edge that does not share an 
end-point with any edge previously placed in the matching.  
The maximal matching oracle essentially emulates this procedure
while selecting a random ranking ``on the fly'', 
but is able to achieve great savings in running time by noting that
to determine whether an edge is placed in the matching, it is only
necessary to know whether or not adjacent edges that are ranked lower
than the current edge have been placed in the matching. 
\ifnum\conf=0
Namely, given an edge $(u,v)$, 
it considers all edges that share an endpoint with $(u,v)$ and whose (randomly
assigned) ranking is 
lower than that of $(u,v)$. If there are no such edges, then the oracle returns {\sc true}.
Otherwise it performs recursive calls to these edges,
where the order of the calls is according to their ranking. If any recursive
call is answered {\sc true}, then the answer on $(u,v)$ is {\sc false}, while
if all answers (on the incident edges with a lower rank) is answered {\sc false},
then the answer on $(u,v)$ is {\sc true}. 
\fi

Though the correctness of the above algorithm follows directly
from the correctness of the greedy algorithm, the query and runtime
analysis are more difficult.
The analysis of \cite{NO} is based on a counting argument
that shows that it is unlikely 
that there is a long path of recursive calls
with a monotone decreasing set of ranks.
Their bound gives a runtime that
is independent of the size of the graph, but exponential in the degree
$d$.   However, using that the algorithm recurses according to the 
smallest ranked neighbor,  
\cite{YYI} give an ingenious analysis that  bounds
by $O(d)$ the total number of expected recursive calls when selecting an edge uniformly
at random, and when selecting a ranking uniformly at random.
This is what allows~\cite{YYI} to obtain an algorithm whose query complexity
and running time are $O(d^4/\eps^2)$.

\ifnum\conf=1
\vspace{-1.5ex}
\fi
\paragraph{Our Algorithm.}
 In what follows we describe an algorithm that has almost linear dependence on
 the maximum degree $d$. The transformation to an algorithm whose complexity 
 depends on the average degree $\bd$ is done on a high level
 along the lines described in~\cite{PR}.
We first depart from {\bf Approx-VC-I} by performing the following variation.
Rather than sampling edges and approximating the size of a maximal matching
by calling the maximal matching oracle on the sampled edges, 
we sample vertices  (as in~\cite{PR}),
and  we call a {\em vertex cover oracle\/} on each selected vertex $v$. 
The vertex cover oracle calls the maximal matching oracle on the edges incident to 
$v$ according to the order induced by their ranking (where the ranking is
selected randomly). Once some edge returns {\sc true},
the vertex cover oracle returns {\sc true}, and if all incident edges return {\sc false},
the vertex cover oracle returns {\sc false}. By performing this variation we
can take a sample of vertices that has size $\Theta(1/\eps^2)$
\ifnum\conf=0
  rather than\footnote{We note that it is actually possible to save one factor of $d$ without
 changing the algorithm {\bf Approx-VC-I} by slightly refining the probabilistic 
 analysis. This would reduce the complexity of {\bf Approx-VC-I} to cubic in $d$.}
 $\Theta(d^2/\eps^2)$. 
\else
and does not depend on any parameter of the graph.
\fi

Unfortunately, the analysis of~\cite{YYI} is no longer applicable
as is. Recall that their analysis bounds the expected number of recursive calls
to the maximal matching oracle, for a random ranking,
and for a {\em randomly selected edge\/}.  In contrast, we select a random 
vertex and call the maximal matching oracle on its (at most $d$) incident edges. 
Nonetheless, we are able to adapt the analysis of~\cite{YYI} and give a
bound of $O(d)$ on the expected number of recursive calls to the maximal matching
oracle, when selecting a vertex uniformly at random.\footnote{To be more precise, we first
give a bound that depends on the ratio between the maximum and minimum degrees
as well as on the average degree, and later we show how to 
obtain a dependence on $d$ (at an extra cost of $1/\eps$)
by slightly modifying the input graph.}

As a direct corollary of the above we can get an algorithm whose query complexity
and running time grow quadratically with $d$. Namely, whenever the maximal matching
oracle is called on a new edge $(u,v)$, the algorithm needs to perform recursive calls on
the edges incident to $u$ and $v$, in an order determined by their ranking. To this
end it can query the $O(d)$ neighbors of $u$ and $v$, assign them (random) rankings,
and continue in a manner consistent with the assigned rankings. 

To reduce the complexity of the algorithm further, we show a method that 
for most of the edges that we visit, allows 
us to query only a small subset of adjacent edges.
Ideally, we would like to make only $k$ queries when 
$k$ recursive calls are made. One of the problems that we encounter here is that
if we do not query all adjacent edges, 
then for some edge $(u,v)$, we could visit 
a different edge incident to $u$ and a different edge 
incident to $v$ and make 
conflicting decisions about the 
ranking of $(u,v)$ from the point of view of these edges. 
This could result in an inconsistent execution of the algorithm with 
results that are hard to predict.
Instead, we devise a probabilistic procedure, that, together with appropriate data structures,
allows us to perform queries almost ``only when needed''
(we elaborate on this  in the next paragraph). 
By this we mean that
we perform queries only on a number of edges that is
a  $\poly(\log (d/\eps))$ factor larger
than the total number of recursive calls made to the maximal matching oracle.
We next discuss our general approach.

As in previous work, we implement the random ranking by assigning 
numbers to edges independently, uniformly at random from $(0,1]$
(or, more precisely, from an appropriate discretization of $(0,1]$).
For each vertex we keep a copy of a data structure that is responsible for
generating and assigning random numbers to incident edges. 
For each vertex, we can ask the corresponding data structure for the incident edge
with  the $i^{\rm th}$ lowest number. 
 How does the data structure work? 
Conceptually, the edges attached to each vertex are grouped into ``layers'',
where the edges in the first layer have random numbers that are at most $1/d$,
the edges in layer $i>1$ have random numbers in the range $2^{i-1}/d$ 
to $2^{i}/d$.   The algorithm randomly chooses 
edges to be in a layer for each vertex, one layer at a time, starting
with the lowest layer.   Each successive
layer is processed only as needed by the algorithm.
If the algorithm decides that an edge is
in the current layer, then it picks a random number for the edge uniformly
from the range associated with the layer.
In particular, it is possible to ensure that the final random number comes from the uniform distribution on $(0,1]$. 
In order to make sure that the same decision is made at both endpoints of an edge $(u,v)$, the data structures for $u$ and $v$ communicate whenever they want to assign a specific random number to the edge. 
The algorithm works in such a way so that vertices need query their incident edges only
when a communication regarding the specific edge occurs.
Our final algorithm is obtained by minimizing the amount of communication between different data structures, and therefore, making them discover not many more edges than necessary for recursive calls in the graph exploration.

\ifnum\conf=1
\vspace{-1.5ex}
\fi
\paragraph{Other Related Work.}
For some restricted classes of graphs it is possible to obtain a 
$(1,\eps)$-estimate of the size of the minimum vertex cover in time that is a function of 
\ifnum\conf=1
only $\eps$~\cite{Elek,HassidimKNO09}.
Similar ideas are used to construct sublinear time estimations of
other parameters of sparse combinatorial objects, such as maximum matching,
set cover, constraint satisfaction \cite{NO, YYI, Y}.
In the related setting of property testing, sublinear time algorithms
are given for testing any class of graphs with
a fixed excluded minor and any property of graphs with a fixed excluded minor \cite{CSS,BSS,Elek,HassidimKNO09,NS}.
\else
only $\eps$.
Elek shows that this is the case for graphs of subexponential growth \cite{Elek}.
 For minor-free graphs, one obtains this result by applying the generic reduction of 
Parnas and Ron~\cite{PR}
to local distributed algorithm of Czygrinow, Ha\'{n}\'{c}kowiak, 
and Wawrzyniak~\cite{CHW08}. 
Both of these results are generalized by Hassidim et al.~\cite{HassidimKNO09}
to any class of hyperfinite graphs. 
In particular, for planar graphs, 
they give
an algorithm that computes a $(1,\eps)$-estimate in $2^{\poly(1/\eps)}$ time. 
While the running time must be exponential in $1/\eps$, 
unless there exists a randomized subexponential algorithm for SAT, 
it remains a neat open question whether the query complexity can be reduced to polynomial in $1/\eps$.

For bipartite graphs, a $(1,\eps n)$-estimate can be computed in $d^{O(1/\eps^2)}$ time. This follows from the relation between the maximum matching size and the minimum vertex size captured by  
K\"onig's theorem and fast approximation algorithms for the maximum matching size~\cite{NO,YYI}.

Ideas similar to those discussed in this paper are used to construct sublinear time estimations of
other parameters of sparse combinatorial objects, such as maximum matching,
set cover, constraint satisfaction \cite{NO, YYI, Y}.
In the related setting of property testing, sublinear time algorithms
are given for testing any class of graphs with
a fixed excluded minor and any property of graphs with a fixed excluded minor \cite{CSS,BSS,Elek,HassidimKNO09,NS}.

\fi
There are also other works on sublinear algorithms for various other graph
measures such as the minimum weight spanning tree~\cite{CRT,CS-mst,CEFMNRS}, 
the average degree~\cite{Feige,GR-avg}, and the number of stars~\cite{GRS}.

%% file: alg.tex
\section{The Oracle-Based Algorithm}
\label{alg.sec}

Let $G = (V,E)$ be an undirected graph with $n$ vertices and $m$ edges, 
where we allow $G$ to contain parallel edges and self-loops.
Let $d$ denote the maximum degree in the graph, and let $\bd$ denote the average degree.
Consider a  {\em ranking\/} $\pi : E \to [m]$ of the edges in
$G = (V,E)$. As noted in the introduction,
such a ranking determines a maximal matching
$\MM^\pi(G)$.
Given $\MM^\pi(G)$, we define a vertex cover $\C^\pi(G)$ as
the set of all endpoints of edges in $\MM^\pi(G)$.
Therefore, $\VCopt \leq |\C^\pi(G)| \leq 2\VCopt$,
where $\VCopt$ is the minimum size of a vertex cover in $G$.
We assume without loss of generality that there are no isolated vertices in
$G$, since such vertices need not belong to any vertex cover.
We shall 
use the shorthand $\MM^\pi$ and $\C^\pi$ for $\MM^\pi(G)$ and
$\C^\pi(G)$, respectively, when $G$ is clear from the context.

\ifnum\conf=0
Assume we have an oracle $\VO^{\pi}$ for a vertex cover based on
a ranking $\pi$ of the edges,
where $\VO^{\pi}(v) = \mbox{\sc true}$ if $v \in \C^\pi(G), \VO^{\pi}(v) = \mbox{\sc false}$ otherwise.
The next lemma follows by applying an additive Chernoff bound.

\BL\label{V-approx.lem}
For any fixed choice of $\pi$, let $\C = \C^\pi(G)$.
Suppose that we uniformly and independently
select $s = \Theta(\frac{1}{\eps^2})$ vertices $v$ from $V$.
Let $t$ be a random variable equal to the number of selected
vertices that belong to $\C$. With high constant probability,
\ifnum\conf=0
$$|\C| - \eps n  \leq \frac{t}{s}\cdot n \leq |\C| + \eps n\;.$$
\else
~$|\C| - \eps n  \leq \frac{t}{s}\cdot n \leq |\C| + \eps n$.
\fi
\EL
\fi

Algorithm~\ref{O_V.alg}, provided below, implements an oracle $\VO^{\pi}$,
that given a vertex $v$, decides whether $v \in \C^\pi$.
This oracle uses another oracle, $\MO^{\pi}$ (described in Algorithm~\ref{alg:O_M})
 that given an edge $e$, decides whether $e \in \MM^\pi$.
Both oracles can determine $\pi(e)$ for any edge $e$ of their choice.
The oracle $\MO^{\pi}$ essentially emulates the greedy algorithm for finding
a maximal matching (based on the ranking $\pi$).
We assume that once the oracle for
the maximal matching decides whether an edge $e$ belongs to
$\MM^\pi$ or not, it records this information in a data structure
that allows to retrieve it later.
\ifnum\conf=0
By Lemma~\ref{V-approx.lem}, 
\else
By Hoeffding's inequality,
\fi
if we perform $\Theta(1/\eps^2)$ calls to
$\VO^{\pi}$, we can get an estimate
of the size of the vertex cover $\C^\pi$
up to an additive error of $(\eps/2)n$,
and hence we can obtain 
a $(2,\eps)$-estimate (as defined in the introduction)
of the size of a minimum vertex cover in $G$.
Hence our focus is on upper bounding the query complexity and
running time of the resulting approximation algorithm when
$\pi$ is selected uniformly at random.

 \ifnum\conf=0
\begin{algorithm}[h] \label{O_V.alg}
Let $e_1,\ldots,e_t$ be the edges incident to the
vertex $v$
in order of increasing rank (that is, $\pi(e_{i+1})> \pi(e_i)$).\\
\For {$i=1,\ldots, t$}{
     \If { $\MO^{\pi}(e_i) =$~{\sc true}}{ \Return {\sc true}}

 \Return {\sc false}
 }
\caption{\small An oracle {\boldmath$\VO^{\pi}(v)$} for a vertex
cover based on a ranking $\pi$ of the edges. Given a vertex $v$, the
oracle returns {\sc true} if $v \in \C^\pi$ and it returns {\sc
false} otherwise.} \label{O_V.fig}
\end{algorithm}
 \else
\begin{algorithm}[h] \label{O_V.alg}
Let $e_1,\ldots,e_t$ be the edges incident to the
vertex $v$ in order of increasing rank
\;
\For{$i=1,\ldots, t$}{
   \lIf{$\MO^{\pi}(e_i) = \mbox{\sc true}$}{\Return {\sc true}}
}
\Return {\sc false}
\caption{\small An oracle {\boldmath$\VO^{\pi}(v)$} for a vertex
cover based on a ranking $\pi$ of the edges. Given a vertex $v$, the
oracle returns {\sc true} if $v \in \C^\pi$ and it returns {\sc
false} otherwise.} \label{O_V.fig}
\end{algorithm}
\fi

\ifnum\conf=0
\begin{algorithm}[h] \label{alg:O_M}
\If {$\MO^{\pi}(e)$ has already been computed}{\Return
the computed answer.}
Let $e_1,\ldots,e_t$ be the edges that share an endpoint with $e$,
in order of increasing rank (that is, $\pi(e_{i+1})> \pi(e_i)$).\;
$i \leftarrow 1$.\;
\While {$\pi(e_i) < \pi(e)$}{
  \If {$\MO^{\pi}(e_i) =$~{\sc true}}{\Return {\sc false}} \Else{ $i \leftarrow i+1$.}
} \Return {\sc true} \caption{ \small  An
oracle 
{\boldmath$\MO^{\pi}(e)$}
for a maximal matching based on a ranking $\pi$ of the edges.
Given an edge $e$, the oracle returns {\sc true} if $e \in \MM^\pi$
and it returns {\sc false} otherwise.} \label{O_M.fig}
\end{algorithm}
\else
\begin{algorithm}[h] \label{alg:O_M}
\lIf{\rm $\MO^{\pi}(e)$ has already been computed}{\Return the computed answer}\;
let $e_1,\ldots,e_t$ be the edges that share an endpoint with $e$,
in order of increasing rank\;
$i := 1$\;
\While{$\pi(e_i) < \pi(e)$}{
\lIf{$\MO^{\pi}(e_i) = \mbox{\sc true}$}{\Return {\sc false}}\;
\lElse{$i \leftarrow i+1$}\;
}
\Return {\sc true}
\caption{ \small  An oracle
{\boldmath$\MO^{\pi}(e)$}
for a maximal matching based on a ranking $\pi$ of the edges.
Given an edge $e$, the oracle returns {\sc true} if $e \in \MM^\pi$
and it returns {\sc false} otherwise.} \label{O_M.fig}
\end{algorithm}
\fi

We start (in Section~\ref{avg-calls.sec})
 by bounding the expected number of calls made to the maximal-matching
oracle $\MO^{\pi}$ in the course of the execution of a call to the vertex-cover
oracle $\VO^{\pi}$.
This bound depends on the average degree in the
graph and on the ratio between the maximum degree and the minimum degree.
A straightforward implementation of the oracles would give
us an upper bound on the complexity of the algorithm that is a factor
of $d$ larger than our final near-optimal algorithm.
In Section~\ref{mm.sec} we
describe a sophisticated method of simulating the behavior of the oracle 
$\MO^\pi$ 
for randomly selected ranking $\pi$, which is
selected ``on the fly''.
Using this method we obtain an algorithm
with only a polylogarithmic overhead (as a function of $d$)
over the number of recursive calls.
Thus, for graphs that are close to being regular, we get an algorithm
whose complexity is $\tilde{O}(d/\eps^2)$.
In Section~\ref{near-opt.sec} we address the issue of
variable degrees, and in particular, show how to get a nearly-linear
dependence on the average degree.

\ifnum\conf=0
\section{Bounding the Expected Number of Calls to the Maximal-Matching Oracle}
\else
\section{Bounding the Number of Calls to the Maximal-Matching Oracle}
\fi
\label{avg-calls.sec}
For a ranking $\pi$ of the edges of a graph
$G$ and a vertex $v\in V$, let $N(\pi,v) = N_G(\pi,v)$ denote the
number of different edges $e$ such that a call $\MO^{\pi}(e)$ was
made to the maximal matching oracle in the course of the
computation of $\VO^{\pi}(v)$. Let $\Pi$ denote the set of all
rankings $\pi$ over the edges of  $G$.
Our goal is to bound the expected value of $N(\pi,v)$ (taken
over a uniformly selected ranking $\pi$ and vertex $v$).
We next state our first main theorem.
\BT\label{avg-calls.thm}
Let $G$ be a graph with $m$ edges and average degree $\bd$, and let the ratio
between the maximum degree and the minimum degree in $G$ be denoted by $\rho$.
The average value of $N(\pi,v)$ taken over all
rankings $\pi$ and vertices $v$ is
$O(\rho\cdot \bd)$.
That is:
\BEQ
\frac{1}{m!}\cdot \frac{1}{n} \cdot \sum_{\pi\in \Pi}\sum_{v\in V}
         N(\pi,v) = O(\rho\cdot \bd) \;.
\EEQ
\ET
If the graph is (close to) regular, then the bound we get in 
Theorem~\ref{avg-calls.thm}  is $O(\bd) = O(d)$. However, for graphs
with varying degrees the bound can be $\Theta(d^2)$. 
As noted previously, we later show how to deal with variable degree graphs
without having to pay a quadratic cost in the maximum degree.

As noted in the introduction,
our analysis builds on the work of Yoshida et al.~\cite{YYI}.
While our analysis does not reduce to theirs\footnote{Indeed,
we initially tried to find such a reduction. The main difficulty we encountered
is that the vertex cover oracle, when executed on a vertex $v$,
 performs calls to the maximal matching
oracle on the edges incident to $v$ until it gets a positive answer (or all the incident
edges return a negative answer). While the analysis of~\cite{YYI} gives us an upper
bound on the expected number
of recursive calls for a given edge, it is not clear how to use such a bound
without incurring an additional multiplicative cost that depends on the degree
of the vertices.}, it uses many of their ideas.
\ifnum\conf=0
\input{alg2.tex}

\else
We next give a high level idea of the proof (where we assume for simplicity
that there are no parallel edges and no self-loops). 
For the full details see
\ifnum\hideappendix=1
the full paper.
\else
Appendix~\ref{avg-calls.app}.
\fi

For any edge $e \in E$, we arbitrarily label its
endpoints by $v_a(e)$ and $v_b(e)$.
For a ranking $\pi$ and an index $k$, let
$\pi_k$ denote the edge $e$ such that $\pi(e) = k$.
We say that an edge $e$ is {\em visited\/}
if a call is made on $e$ either in the course of an oracle computation of
$\VO^{\pi}(v_a(e))$ or $\VO^{\pi}(v_b(e))$,
or in the
course of an oracle computation of $\MO^{\pi}(e')$ for an
 edge $e'$ that shares an endpoint with $e$.
For a vertex $v$ and an edge $e$, let $X^\pi(v,e) = X^\pi_G(v,e)$
equal $1$ if $e$ is visited in the course of the execution of
$\VO^{\pi}(v)$.
Using the notation just introduced, we have that
$N(\pi,v) \;=\; \sum_{e\in E} X^\pi(v,e)$.

Turning to the ``point of view'' of an edge $e$, for each rank $k$, we let
 ~$X_k(e) \eqdef \sum_{\pi\in \Pi} \Big( X^\pi(v_a(\pi_k),e) + X^\pi(v_b(\pi_k),e) \Big)$.
That is, $X_k(e)$ is the total number of calls made to the
maximal matching oracle on the edge $e$ when summing over all
rankings $\pi$, and performing an oracle call to the vertex-cover
oracle from one of the endpoints of $\pi_k$. Observe that
$\sum_{k=1}^m X_k(e) \;=\; \sum_{\pi\in \Pi}
\sum_{v\in V}
               {\rm deg}(v) \cdot X^\pi(v,e)$,
where ${\rm deg}(v)$ denotes the degree of $v$ in the graph.
Thus, if we can upper-bound $X_k(e)$, then we can get 
an upper bound on the sum over all $\pi$ and $v$ of $N(\pi,v)$
(as in Theorem~\ref{avg-calls.thm}).

Indeed, we prove that $X_k(e) \leq 2(m-1)! + (k-1)\cdot (m-2)! \cdot d$.
This upper bound on $X_k(e)$ is obtained by induction, where 
at the heart of the induction step is the
bound $X_{k+1}(e)  - X_k(e) \leq (m-2)! \cdot d$.
In order to obtain this bound on the difference between $X_{k+1}(e)  $
and $X_k(e)$ we adapt the following idea used in~\cite{YYI} to our
needs. For each ranking $\pi$, we consider a small ``mutation'' of
$\pi$, in which the edges with ranks $k$ and $k+1$ ``switch ranks'', while
all other edges remain with the same rank. Using this idea we can get
a bound on  $X_{k+1}(e)  - X_k(e)$ that only depends on rankings (and their
``mutations'') in which these two edges share a common endpoint. By a 
(slightly tedious) case analysis we can further restrict the relevant rankings that
contribute to $X_{k+1}(e)  - X_k(e)$. In the final step we adapt another
clever idea from~\cite{YYI}, where a one-to-one mapping between sets of rankings
is presented, and the analysis is completed.

\fi

%% file: alg2.tex
We start by making a very simple but useful observation about the maximal
matching oracle $\MO^{\pi}$ (Algorithm~\ref{alg:O_M}), which follows immediately
from the definition of the oracle.
\begin{obser}\label{ranksGoDown.obs}
For any edge $e$, consider the execution of $\MO^{\pi}$ on $e$. If in the course
of this execution, a recursive call is made to $\MO^{\pi}$ on another edge $e'$,
then necessarily $\pi(e') < \pi(e)$.
Therefore, for any consecutive sequence of (recursive) calls to edges $e_\ell,\ldots,e_1$,
$\pi(e_\ell) > \pi(e_{\ell-1}) >  \ldots > \pi(e_1)$.
\end{obser}
\ifnum\conf=0
In order to prove Theorem~\ref{avg-calls.thm} we introduce
more notation. 
\else
For the sake of completeness, we reintroduce the notation presented
in Section~\ref{avg-calls.sec}, and restate Theorem~\ref{avg-calls.thm}.
For a ranking $\pi$ of the edges of a graph
$G$ and a vertex $v\in V$, let $N(\pi,v) = N_G(\pi,v)$ denote the
number of different edges $e$ such that a call $\MO^{\pi}(e)$ was
made to the maximal matching oracle in the course of the
computation of $\VO^{\pi}(v)$. Let $\Pi$ denote the set of all
rankings $\pi$ over the edges of  $G$.
Our goal is to bound the expected value of $N(\pi,v)$ (taken
over a uniformly selected ranking $\pi$ and vertex $v$).

\medskip\noindent{\bf Theorem~\ref{avg-calls.thm} (Restated)}~
{\it
Let $G$ be a graph with $m$ edges and average degree $\bd$, and let the ratio
between the maximum degree and the minimum degree in $G$ be denoted by $\rho$.
The average value of $N(\pi,v)$ taken over all
rankings $\pi$ and vertices $v$ is
$O(\rho\cdot \bd)$.
That is:
\BEQ
\frac{1}{m!}\cdot \frac{1}{n} \cdot \sum_{\pi\in \Pi}\sum_{v\in V}
         N(\pi,v) = O(\rho\cdot \bd) \;.
\EEQ
}

\medskip
\fi
For any edge $e \in E$, we arbitrarily label its
endpoints by $v_a(e)$ and $v_b(e)$ (where if $e$ is a self-loop then
$v_a(e) = v_b(e)$, and if $e$ and $e'$ are parallel edges, then
$v_a(e) = v_a(e')$ and $v_b(e) = v_b(e')$).
For a ranking $\pi$ and an index $k$, let
$\pi_k$ denote the edge $e$ such that $\pi(e) = k$.

We say that an edge $e$ is {\em visited\/}
if a call is made on $e$ either in the course of an oracle computation of
$\VO^{\pi}(v_a(e))$ or $\VO^{\pi}(v_b(e))$ (that is, as a non-recursive call), or in the
course of an oracle computation of $\MO^{\pi}(e')$ for an
 edge $e'$ that shares an endpoint with $e$ (as a
recursive call).
For a vertex $v$ and an edge $e$, let $X^\pi(v,e) = X^\pi_G(v,e)$
equal $1$ if $e$ is visited in the course of the execution of
$\VO^{\pi}(v)$.
Using the notation just introduced, we have that
\BEQ\label{Npiv-Xpiv.eq}
N(\pi,v) \;=\; \sum_{e\in E} X^\pi(v,e)\;.
\EEQ

\begin{obser}\label{smallerRankFalse.obs}
Let $e = (v,u)$.
If $X^\pi(v,e) = 1$, then for each edge $e'$ that shares
the endpoint $v$ with $e$ and for which $\pi(e') < \pi(e)$ we have that
$\MO^{\pi}(e') = \mbox{\sc false}$.
\end{obser}
\sloppy
To verify Observation~\ref{smallerRankFalse.obs},
assume, contrary to the claim, that there
 exists an edge $e'$  as described
in the observation and $\MO^{\pi}(e') = \mbox{\sc true}$.
We first note that by Observation~\ref{ranksGoDown.obs},
the edge $e$ cannot be visited in the course of an execution of
$\MO^{\pi}$ on any edge $e'' = (v,w)$ such that $\pi(e'') < \pi(e)$
(and in particular this is true for $e'' = e'$). Since $\VO^{\pi}(v)$
performs calls to the edges incident to $v$ in order of increasing rank,
if $\MO^{\pi}(e') = \mbox{\sc true}$, then $\VO^{\pi}(v)$ returns
{\sc true} without making a call to $\MO^{\pi}(e)$.
This contradicts the premise of the observation that $X^\pi(v,e) = 1$.

The next notation is central to our analysis. For $k \in [m]$ and a
fixed edge $e$:
\BEQ\label{Xk-def.eq}
 X_k(e) \eqdef \sum_{\pi\in \Pi} \Big( X^\pi(v_a(\pi_k),e) + X^\pi(v_b(\pi_k),e) \Big)\;.
\EEQ
That is, $X_k(e)$ is the total number of calls made to the
maximal matching oracle on the edge $e$ when summing over all
rankings $\pi$, and performing an oracle call to the vertex-cover
oracle from one of the endpoints of $\pi_k$. Observe that
\BEQ\label{sumXk.eq}
\sum_{k=1}^m X_k(e) \;=\; \sum_{\pi\in \Pi}
\sum_{v\in V}
               {\rm deg}(v) \cdot X^\pi(v,e)
\EEQ 
where ${\rm deg}(v)$ denotes the degree of $v$ in the graph,
and for simplicity of the presenation we count each self-loop
as contributing $2$ to the degree of the vertex.
We next give an upper bound on $X_k(e)$.
\BL\label{Xk.lem}
For every edge $e$ and every  $k \in [m]$:
\BEQ
X_k(e) \leq 2(m-1)! + (k-1)\cdot (m-2)! \cdot d\;. \EEQ
\EL
In order to prove Lemma~\ref{Xk.lem}, we establish the following lemma.

\BL\label{Xk-def.lem}
For every edge $e$ and every  $k \in [m-1]$:
\BEQ
X_{k+1}(e)  - X_k(e) \leq (m-2)! \cdot d\;. \EEQ
\EL
Before proving Lemma~\ref{Xk-def.lem}, we show how Lemma~\ref{Xk.lem} easily
follows from it.

\BPFOF{Lemma~\ref{Xk.lem}} We prove the lemma by induction on $k$.
For the base case, $k=1$,
\BEQ
  X_1(e) = \sum_\pi \Big( X^\pi(v_a(\pi_{1}),e) +
              X^\pi(v_b(\pi_{1}),e)\Big)\;.
\EEQ
By the definition of the vertex-cover oracle, when starting
from either $v_a(\pi_1)$ or from $v_b(\pi_1)$, only a single call is
made to the maximal matching oracle. This call is on the
edge $\pi_1$, which returns {\sc true} without making
any further calls, because all edges (that share an
endpoint with $\pi_1$) have a larger rank. This implies that if $e =
\pi_1$, then $X^\pi(v_a(\pi_{1}),e) = X^\pi(v_b(\pi_{1}),e) = 1$, and
otherwise $X^\pi(v_a(\pi_{1}),e) = X^\pi(v_b(\pi_{1}),e) =0$. For
any fixed $e$, the number of rankings $\pi$ such that $e = \pi_1$ is simply $(m-1)!$ and so
$X_1(e) = 2(m-1)!$, as required.

We now turn to the induction step. Assuming the induction hypothesis
holds for $k-1 \geq 1 $, we prove it for $k > 1$. This follows
directly from Lemma~\ref{Xk-def.lem} (and the induction hypothesis):
\BEQN
X_k(e) &\leq& X_{k-1}(e) + (m-2)! \cdot d \\
&\leq&  2(m-1)! + (k-2)\cdot (m-2)! \cdot d + (m-2)! \cdot d \\
&=& 2(m-1)! + (k-1)\cdot (m-2)! \cdot d\;,
\EEQN
and the lemma is established.
\EPFOF

\medskip
\BPFOF{Lemma~\ref{Xk-def.lem}} Throughout the proof we fix $k$ and
$e$. For a ranking $\pi$, let $\pi'$ be defined as follows:
$\pi'_{k+1} = \pi_k$, $\pi'_k = \pi_{k+1}$ and $\pi'_j = \pi_j$ for
every $j \notin \{k,k+1\}$.
\begin{obser} \label{lessThanKSameRun.obs}
If $\pi$ and $\pi'$ are as defined above,
then for each edge $e$ where $\pi(e) < k$ (and therefore, $\pi'(e) < k$):
$\MO^{\pi}(e) = \MO^{\pi'}(e) $.
\end{obser}
Observation~\ref{lessThanKSameRun.obs} is true due to the fact
that if $\pi(e) < k$ then by the definition of $\pi'$, we have that
$\pi'(e) = \pi(e)$. Since in a recursive call we only go to an edge with a lower rank
(see Observation~\ref{ranksGoDown.obs}), we get that
the execution of $\MO^{\pi}(e)$ is equivalent to the execution of $\MO^{\pi'}(e)$.

We shall use the notation $\Pi_k$ for
those rankings $\pi$ in which $\pi_k$ and $\pi_{k+1}$ share a common
endpoint. Note that if $\pi \in \Pi_k$, then $\pi' \in \Pi_k$ as
well (and if $\pi \notin \Pi_k$, then $\pi' \notin \Pi_k$).
For two edges $e = (v_1,v_2)$ and $e' = (v_2, v_3)$ (which share a
common endpoint $v_2$), we let $v_c(e,e') = v_c(e',e) = v_2$ (`$c$' for
`common') and $v_d(e,e') = v_1$, $v_d(e',e) = v_3$ (`$d$' for
`different').
If $e$ and $e'$ are parallel edges, then
we let $v_d(e,e') = v_d(e',e)$ be $v_a(e) = v_a(e')$
and $v_c(e,e') = v_c(e',e)$ be $v_b(e) = v_b(e')$.
If $e$ is a self-loop on a vertex $v_1$ that is also
an endpoint of
 $e'$ (so that $v_2 = v_1$),  then $v_d(e,e') = v_c(e,e') = v_1$.

\noindent
For any edge $e$ and for $1 \leq k \leq m-1$, 
\BEQN
\lefteqn{X_{k+1}(e) - X_k(e)}\nonumber \\
&=& \sum_\pi \Big(X^\pi(v_a(\pi_{k+1}),e) + X^\pi(v_b(\pi_{k+1}),e)\Big)
   - \sum_\pi \Big(X^\pi(v_a(\pi_k),e) + X^\pi(v_b(\pi_k),e)\Big) \\
&=& \sum_{\pi \in \Pi_k} \Big(X^\pi(v_a(\pi_{k+1}),e) + X^\pi(v_b(\pi_{k+1}),e)\Big)
   - \sum_{\pi \in\Pi_k} \Big(X^\pi(v_a(\pi_k),e) + X^\pi(v_b(\pi_k),e)\Big) \nonumber\\
&&+\; \sum_{\pi \notin \Pi_k}\Big(X^\pi(v_a(\pi_{k+1}),e) + X^\pi(v_b(\pi_{k+1}),e)\Big)
   - \sum_{\pi \notin\Pi_k} \Big(X^\pi(v_a(\pi_k),e) + X^\pi(v_b(\pi_k),e)\Big)  \\
&=& \sum_{\pi \in \Pi_k} X^\pi(v_c(\pi_{k+1},\pi_{k}),e)
   - \sum_{\pi \in\Pi_k}  X^\pi(v_c(\pi_k,\pi_{k+1}),e) \label{pi-in-Pk-c.eq}\\
&&+\; \sum_{\pi \in \Pi_k} X^\pi(v_d(\pi_{k+1},\pi_{k}),e)
   - \sum_{\pi \in\Pi_k}  X^\pi(v_d(\pi_k,\pi_{k+1}),e) \label{pi-in-Pk-d.eq}\\
&&+\; \sum_{\pi \notin \Pi_k} \Big(X^\pi(v_a(\pi_{k+1}),e) + X^\pi(v_b(\pi_{k+1}),e)\Big)
   - \sum_{\pi \notin\Pi_k} \Big(X^\pi(v_a(\pi_k),e) + X^\pi(v_b(\pi_k),e)\Big)
                      \label{pi-notin-Pk.eq} . 
\EEQN
By the definition of $v_c(\cdot,\cdot)$, for every $\pi \in \Pi_k$ we have that
$v_c(\pi_{k+1},\pi_k) = v_c(\pi_k,\pi_{k+1})$ and so
\BEQ
X^\pi(v_c(\pi_{k+1},\pi_k),e) =
X^\pi(v_c(\pi_k,\pi_{k+1}),e) \label{vcEqual} \;,
\EEQ
implying that the
expression in Equation~(\ref{pi-in-Pk-c.eq}) evaluates to $0$. Since
$\pi' \in \Pi_k$ if and only if $\pi \in \Pi_k$, we get that
\BEQ
 \sum_{\pi \in \Pi_k} X^\pi(v_d(\pi_{k+1},\pi_{k}),e)
  = \sum_{\pi' \in \Pi_k} X^{\pi'}(v_d(\pi'_{k+1},\pi'_{k}),e)
  = \sum_{\pi \in \Pi_k} X^{\pi'}(v_d(\pi'_{k+1},\pi'_{k}),e)\;,
\label{vdEqual} \EEQ
and
\BEQN
\sum_{\pi \notin \Pi_k} \Big(X^\pi(v_a(\pi_{k+1}),e) +X^\pi(v_b(\pi_{k+1}),e)] = \sum_{\pi' \notin \Pi_k}
\Big(X^{\pi'}(v_a(\pi'_{k+1}),e) + X^{\pi'}(v_b(\pi'_{k+1}),e)\Big) \nonumber \\
= \sum_{\pi \notin \Pi_k} \Big(X^{\pi'}(v_a(\pi'_{k+1}),e) + X^{\pi'}(v_b(\pi'_{k+1}),e)\Big)\;.
\label{vjNotInKEqual} \EEQN
Therefore,
\BEQN X_{k+1}(e) - X_k(e)
&=& \sum_{\pi \in \Pi_k} X^{\pi'}(v_d(\pi'_{k+1},\pi'_{k}),e)
    - \sum_{\pi \in\Pi_k}  X^\pi(v_d(\pi_k,\pi_{k+1}),e) \nonumber \\
&& +\; \sum_{\pi \notin \Pi_k}\Big(X^{\pi'}(v_a(\pi'_{k+1}),e) + X^{\pi'}(v_b(\pi'_{k+1}),e)\Big)\nonumber \\
&& -\; \sum_{\pi \notin\Pi_k} \Big(X^\pi(v_a(\pi_k),e) + X^\pi(v_b(\pi_k),e)\Big)\;.
\label{Xk1-Xk-2.eq} \EEQN
The next useful observation is that for every $\pi \notin \Pi_k$
(and for every $e$ and $j \in \{a, b\}$),
\BEQ\label{pi-not-in-Pk-2.eq}
 X^{\pi'}(v_j(\pi'_{k+1}),e) = X^\pi(v_j(\pi_k),e)\;.
\EEQ
This follows by combining the fact that
$v_j(\pi'_{k+1}) = v_j(\pi_k)$ with Observations~\ref{ranksGoDown.obs}
and~\ref{lessThanKSameRun.obs}.

By combining Equation~(\ref{Xk1-Xk-2.eq}) with
Equation~(\ref{pi-not-in-Pk-2.eq}) we obtain that
\BEQ
X_{k+1}(e) - X_k(e)
\;=\; \sum_{\pi \in \Pi_k} X^{\pi'}(v_d(\pi'_{k+1},\pi'_{k}),e)
    - \sum_{\pi \in\Pi_k}  X^\pi(v_d(\pi_k,\pi_{k+1}),e)\;.
\EEQ

\begin{figure*}[htb]
\begin{center}
\includegraphics{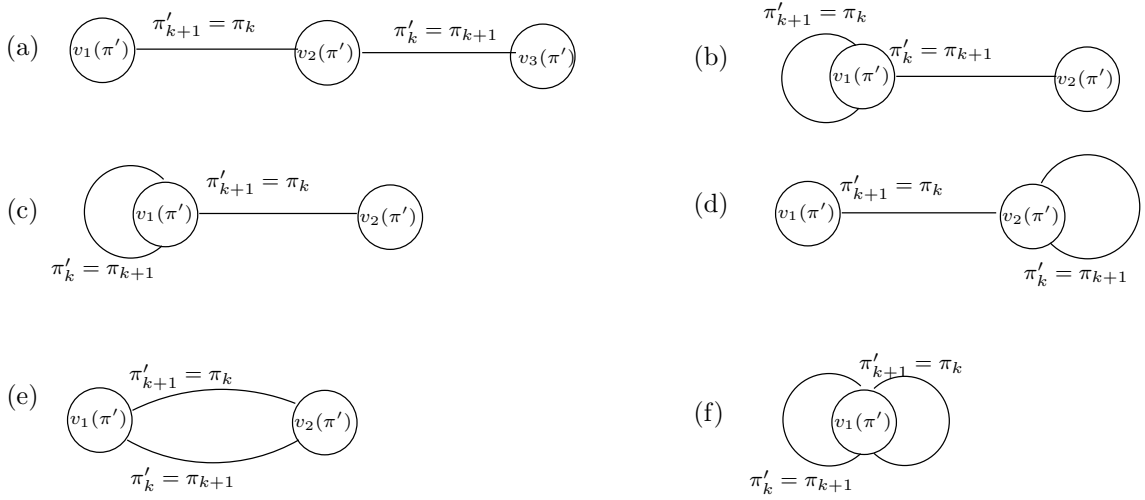}
\end{center}
\caption[Short Caption]{\small An illustration for the various cases
in which $\pi \in \Pi_k$ (i.e., $\pi'_k$ and $\pi'_{k+1}$ share at least one endpoint) 
and we need to compare the executions of
$\VO^{\pi}(v_1(\pi'))$ and $\VO^{{\pi'}}(v_1(\pi'))$ (where $v_1(\pi') =
v_d(\pi'_{k+1},\pi'_{k}) = v_d(\pi_k,\pi_{k+1})$). We refer to the different cases (a)--(f)
in the analysis.}
\label{OnlyCase.fig}
\end{figure*}

Therefore, we need only consider executions in which the underlying
rankings $\pi$ and $\pi'$ belong to $\Pi_k$, and the execution starts
from the vertex  $v_1(\pi') = v_d(\pi'_{k+1},\pi'_{k}) =
v_d(\pi_k,\pi_{k+1})$.
We shall use the shorthand notation
$v_2(\pi') = v_c(\pi'_{k+1},\pi'_{k}) = v_c(\pi_k,\pi_{k+1})$, and $v_3(\pi') =
v_d(\pi'_{k},\pi'_{k+1})  =  v_d(\pi_{k+1},\pi_{k})$.
For an illustration, see Figure~\ref{OnlyCase.fig}.
We shall make use of the following simple observation.
\begin{obser} \label{selfLoopReturnTrue.obs}
Let $e$ be a self-loop. For any vertex $v$ and ranking $\pi$, 
if in the course of the execution of $\VO^\pi(v)$ 
a call is made to $\MO^{\pi}(e)$, then $\MO^{\pi}(e)=\mbox{\sc true}$.
\end{obser}
Observation~\ref{selfLoopReturnTrue.obs} is true since if a call 
is made to $\MO^{\pi}(e)$ where $e$ is a self-loop, i.e., $e = (v,v)$ for some vertex $v$, 
then from Observation~\ref{smallerRankFalse.obs} we know that all other edges incident to $v$ 
with ranks smaller than $\pi(e)$ return {\sc false}. 
Therefore, by the definition of $\MO^{\pi}$ we get that 
$\MO^{\pi}(e)=\mbox{\sc true}$.

\begin{figure*}[htb]
\begin{center}
\includegraphics{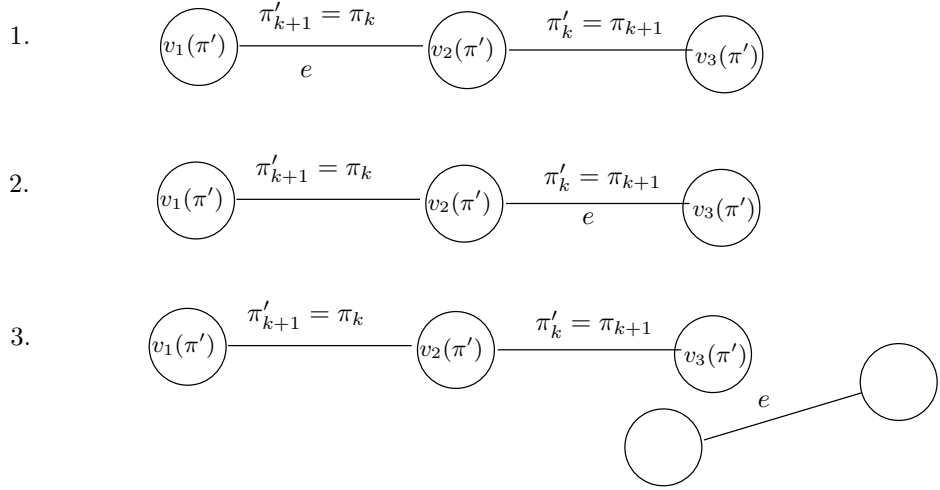}
\end{center}
\caption{\small An illustration for the three 
possible (sub-)cases when $\pi' \in \Pi_k$: 
1. $e = (v_1(\pi'),v_2(\pi'))$;    
2. $e = (v_2(\pi'),v_3(\pi'))$;    
3. $e \notin \{(v_1(\pi'),v_2(\pi')),(v_2(\pi'),v_3(\pi'))\}$. 
This illustration corresponds to Case (a) in  Figure~\ref{OnlyCase.fig}
(i.e., no self-loops and no parallel edges).}
\label{ThreePoss.fig}
\end{figure*}

We would like to understand when $X^{\pi'}(v_1(\pi'),e)= 1$ 
while $X^{\pi}(v_1(\pi'),e)= 0$. We consider three possible cases
(for an illustration see Figure~\ref{ThreePoss.fig}) :
\BE
\item $e = (v_1(\pi'),v_2(\pi'))$ (so that  $\pi'(e) = k+1$ and  $\pi(e) = k$).
In this case, if $X^{\pi'}(v_1(\pi'),e)= 1$, then $X^{\pi}(v_1(\pi'),e)= 1$.
To verify this, note that if $X^{\pi'}(v_1(\pi'),e)= 1$ then by
Observation~\ref{smallerRankFalse.obs},
$\MO^{\pi'}(e')=\mbox{\sc false}$ for each edge $e'$ where $v_1(\pi')$
is one of its endpoints and $\pi'(e') < k+1$.
By applying Observation~\ref{lessThanKSameRun.obs} we get that for each
edge $e'$ such that $\pi(e') < k$ we have that $\MO^{\pi}(e') = \MO^{\pi'}(e')$.
Therefore,  for each edge $e'$ such that $\pi(e') < k$ and
$v_1(\pi')$ is one of its endpoints we have that
 $\MO^{\pi}(e') = \MO^{\pi'}(e') = \mbox{\sc false}$.
Hence $X^{\pi}(v_1(\pi'),e)=1$.

We note that if $\pi'_k$ is a self-loop (see cases (c) and (f) in  Figure~\ref{OnlyCase.fig}),
then by Observation~\ref{selfLoopReturnTrue.obs} we have that $\MO^{\pi'}(\pi'_k) = \mbox{\sc true}$.
By the definition of $\VO^{\pi'}$ this implies that $\pi'_{k+1} = e$ will not be visited 
in the course of the execution of $\VO^{\pi'}(v_1(\pi'))$, so that $X^{\pi'}(v_1(\pi'),e)$ is necessarily $0$. 

\item $e = (v_2(\pi'),v_3(\pi'))$,
(so that $\pi(e) = k+1$ and $\pi'(e) = k$). In this case it is
possible (though not necessary) that $X^{\pi'}(v_1(\pi'),e)= 1$ and
$X^{\pi}(v_1(\pi'),e)= 0$.

\item $e \notin \{(v_1(\pi'),v_2(\pi')),(v_2(\pi'),v_3(\pi'))\}$.
In this case it is also possible (though not necessary) that
$X^{\pi'}(v_1(\pi'),e)= 1$ and $X^{\pi}(v_1(\pi'),e)= 0$. 

Out of all cases illustrated
in Figure~\ref{OnlyCase.fig}, this is possible only in cases (a) and (b). 
We next explain why it is not possible in all other cases.
\BI
  \item Case (c). If $\VO^{\pi'}(v_1(\pi'))$ visits $e$ before it visits $\pi'_{k}$, 
then so does $\VO^{\pi}(v_1(\pi'))$ (from Observation~\ref{lessThanKSameRun.obs}). 
Otherwise, $\VO^{\pi'}(v_1(\pi'))$  visits $\pi'_{k}$ first, but since it is a self-loop, 
from Observation~\ref{selfLoopReturnTrue.obs} we have that $\MO^{\pi'}(\pi'_k) = \mbox{\sc true}$. 
 By the definition of $\VO^{\pi'}$ we get that $X^{\pi'}(v_1(\pi'),e) = 0$.

 \item Case (d). If $\VO^{\pi'}(v_1(\pi'))$ visits $e$ before it visits $\pi'_{k+1}$, 
then so does $\VO^{\pi}(v_1(\pi'))$ (from Observation~\ref{lessThanKSameRun.obs}). 
Otherwise, if $\VO^{\pi'}(v_1(\pi'))$ visits $\pi'_{k+1}$ and $e$ in the same sequence of
recursive calls without visiting $\pi'_{k}$, then so does $\VO^{\pi}(v_1(\pi'))$. 
If there is no such sequence, then $\VO^{\pi'}(v_1(\pi'))$ will visit 
$\pi'_{k+1}$ and $\pi'_{k}$. Since $\pi'_{k}$ is a self-loop, 
from Observation~\ref{selfLoopReturnTrue.obs} we have that
$\MO^{\pi'}(\pi'_k) = \mbox{\sc true}$, implying that $\MO^{\pi'}(\pi'_{k+1}) = {\sc false}$. 
Therefore, the sequence of recursive calls that visits $e$ in
the execution of $\VO^{\pi'}(v_1(\pi'))$, starts from
an edge incident to $v_1(\pi')$ whose rank is  greater than $k+1$,
and the same sequence of calls is made in the execution of $\VO^{\pi}(v_1(\pi'))$.
  \item Case (e). Since the edges are parallel, 
if there is a sequence of recursive calls that  visits $e$ in
the execution of $\VO^{\pi'}(v_1(\pi'))$, then there is such a sequence
in the  execution of $\VO^{\pi}(v_1(\pi'))$, where the only difference is
that the first sequence includes $\pi'_k$ while the second includes $\pi_k$
(which are parallel edges).

  \item Case (f). If $\VO^{\pi'}(v_1(\pi'))$ visits $e$ in a sequence of
recursive calls that starts with an edge having rank smaller than $k$,
 then from Observation~\ref{lessThanKSameRun.obs} so will $\VO^{\pi}(v_1(\pi'))$. 
Otherwise, since $\pi'_k$ is a self-loop, by Observation~\ref{selfLoopReturnTrue.obs}, if a call
is made to  $\MO^{\pi'}(\pi'_k)$, then it returns {\sc true}, causing 
the execution of $\VO^{\pi'}(v_1(\pi'))$ to terminate without visiting any
additional edges (so that $e$ cannot be visited in a sequence of 
recursive calls that starts with an edge having rank at least $k$). 
\EI
\EE

For a fixed edge $e$
we shall use the following notation for the sets of
rankings that correspond to the last two cases described above.
Specifically:
\BI
\item Let $\Pi^{e,1} = \Pi^{e,1}_k$ denote the set of all rankings $\pi' \in \Pi_k$
where $e = (v_2(\pi'),v_3(\pi'))$ and $X^{\pi'}(v_1(\pi'),e)= 1$. (Here we
shall make the worst case assumption that $X^{\pi}(v_1(\pi'),e)= 0$).
\item Let $\Pi^{\neg e} = \Pi^{\neg e}_k$ denote the set of all rankings  $\pi' \in \Pi_k$
where $e \notin \{(v_1(\pi'),v_2(\pi')),(v_2(\pi'),v_3(\pi'))\}$ and
$X^{\pi'}(v_1(\pi'),e)= 1$ while $X^{\pi}(v_1(\pi'),e)= 0$.
\EI
Thus,
$X_{k+1}(e) - X_k(e) \leq |\Pi^{e,1}|+|\Pi^{\neg e}|$. In order to upper bound
$|\Pi^{e,1}|+|\Pi^{\neg e}|$, we consider another set of rankings: \BI
\item Let $\Pi^{e,0} = \Pi^{e,0}_k$ denote the set of all rankings $\pi' \in \Pi_k$
such that $e = (v_2(\pi'),v_3(\pi'))$ and $X^{\pi'}(v_1(\pi'),e)= 0$. \EI
By the definition of $\Pi^{e,1}$ and $\Pi^{e,0}$, we have that 
\BEQ
|\Pi^{e,1}| +|\Pi^{e,0}| \leq (m-2)!\cdot d\;. 
\EEQ
 This is true since each ranking $\pi' \in \Pi^{e,1}\cup \Pi^{e,0}$ is determined
by first setting $\pi'(e) = k$, then selecting another edge incident
to the endpoint $v_2(\pi')$ of $e$ 
(if $e$ is a self-loop then $v_2(\pi') = v_1(\pi')$) 
and giving it rank $k+1$ 
(where there are at most $\deg(v_2(\pi'))-1 \leq d-1$
such edges), and finally selecting one of the possible $(m-2)!$
rankings for the remaining edges. We next prove that $|\Pi^{\neg e}| \leq |\Pi^{e,0}|$,
from which Lemma~\ref{Xk-def.lem} follows.

\noindent To this end we prove the next claim.
\BCM\label{inject.clm}
There is an injection from $\Pi^{\neg e}$ to $\Pi^{e,0}$.
\ECM
The proof of Claim~\ref{inject.clm} is very similar to a proof of
a corresponding claim in~\cite{YYI}, but due to our need to extend the proof to a graph 
with self-loops and parallel edges, and also due to several additional differences, we
include it here for completeness.

\BPF
We start by making the following observations:
\begin{obser}\label{onlyPossiblePathA.obs}
If $\pi' \in \Pi^{\neg e}$ and we are in Case (a) as illustrated in
Figure~\ref{OnlyCase.fig}, then in the course of the
 execution of $\VO^{\pi'}(v_1(\pi'))$
there is a consecutive sequence of recursive calls that includes $\pi'_{k+1}$, $\pi'_k$ and $e$ at the end.
That is, there is a sequence of recursive calls corresponding
to a  path of edges $(e_\ell, e_{\ell-1} \ldots e_1)$ such that
$e_\ell = \pi'_{k+1}, e_{\ell-1} = \pi'_k$ and $e_1 = e$.
\end{obser}
\sloppy
To verify Observation~\ref{onlyPossiblePathA.obs}, note that
since $\pi' \in \Pi^{\neg e}$ we know that
$X^{\pi'}(v_1(\pi'),e)= 1$ and $X^{\pi}(v_1(\pi'),e)= 0$.
The only difference between the execution of
$\VO^{\pi'}(v_1(\pi'))$ and $\VO^{\pi}(v_1(\pi'))$
is that $\MO^{\pi'}(\pi'_{k+1})$ can call
$\MO^{\pi'}(\pi'_{k})$ but $\MO^{\pi}(\pi'_{k+1}) = \MO^{\pi}(\pi_{k})$
cannot call $\MO^{\pi}(\pi'_{k}) = \MO^{\pi}(\pi_{k+1})$.
Thus, the only way that $\VO^{\pi'}(v_1(\pi'))$ and $\VO^{\pi}(v_1(\pi'))$ will
create different sequences of recursive calls is when
$\VO^{\pi'}(v_1(\pi'))$ calls $\MO^{\pi'}(\pi'_{k+1})$
and then $\MO^{\pi'}(\pi'_{k+1})$ calls $\MO^{\pi'}(\pi'_{k})$.
Furthermore, these two calls have to be one after the other,
since by Observation~\ref{ranksGoDown.obs}, the ranks can only decrease in
a sequence of recursive calls.

\begin{obser}\label{onlyPossiblePathB.obs}
If $\pi' \in \Pi^{\neg e}$ and we are in Case (b) as illustrated in Figure~\ref{OnlyCase.fig}, 
then in the course of the execution of $\VO^{\pi'}(v_1(\pi'))$
there is a consecutive sequence of recursive calls that starts with $\pi'_k$,
and ends with $e$ (so that, in particular, it does not include $\pi'_{k+1}$).
That is, there is a sequence of recursive calls corresponding
to a  path of edges $( e_{\ell-1} \ldots e_1)$ such that
$e_{\ell-1} = \pi'_k$ and $e_1 = e$.
\end{obser}
To verify Observation~\ref{onlyPossiblePathB.obs}, note that
since $\pi' \in \Pi^{\neg e}$ we know that
$X^{\pi'}(v_1(\pi'),e)= 1$ and $X^{\pi}(v_1(\pi'),e)= 0$. 
The execution of $\VO^{\pi'}(v_1(\pi'))$ cannot visit $e$ in the course of 
a sequence of recursive calls starting from an edge incident to $v_1(\pi')$ 
where the edge has ranking smaller $k$. Otherwise, from Observation~\ref{lessThanKSameRun.obs} 
 we would get that $\VO^{\pi}(v_1(\pi'))$ also visits $e$ 
which contradicts the premise that $\pi' \in \Pi^{\neg e}$.
We also know that $\VO^{\pi'}(v_1(\pi'))$ cannot visit $\pi'_{k+1}$. 
If it would have, then since it is a self-loop, from Observation~\ref{selfLoopReturnTrue.obs}, 
$\MO^{\pi'}(\pi'_{k+1}) = \mbox{\sc true}$, 
causing $\VO^{\pi'}(v_1(\pi'))$ to terminate without visiting $e$, 
which contradicts $X^{\pi'}(v_1(\pi'),e)= 1$. 

\medskip
We shall now prove Claim~\ref{inject.clm}. 
Let $\pi^1$ be a ranking in  $\Pi^{\neg e}$ (so that  $\pi^1(e) \notin \{k,k+1\}$).
By the definition of $\Pi^{\neg e}$ and by Observations~\ref{onlyPossiblePathA.obs} 
and~\ref{onlyPossiblePathB.obs}, we have the following.
In Case (a), the execution of $\VO^{{\pi^1}}(v_1(\pi^1))$ induces a sequence of (recursive) calls
to the maximal matching oracle, where this sequence corresponds to a
path $P = (e_\ell, \ldots , e_1)$ such that $e_\ell = \pi^1_{k+1}$,
$e_{\ell-1} = \pi^1_k$, and $e_1 = e$.
In Case (b), the execution of $\VO^{{\pi^1}}(v_1(\pi^1))$ induces a sequence of (recursive) calls
to the maximal matching oracle, where this sequence corresponds to a
path $P' = (e_{\ell-1}, \ldots , e_1)$ such that 
$e_{\ell-1} = \pi^1_k$, and $e_1 = e$.
Since in Case (b) $P$ is also a path in the graph, we may refer to the path $P$
in both cases (and take into account, if needed, that in Case (b) $e_\ell=\pi^1_{k-1}$
is a self-loop and is not part of the sequence of recursive calls that reaches $e$).
While we do not know the rankings of the edges $e_{\ell-2},\ldots e_1$, we know 
from Observation~\ref{ranksGoDown.obs} that they are in monotonically decreasing order, 
and that they are all smaller than $k$.
We also know that the path does not include any parallel edges. 
This is true since if $e_t$ and $e_{t-1}$ are adjacent edges in the path $P$ 
and they are parallel edges, then 
from Observation~\ref{lowerRanks.obs} $\pi'(e_{t-1}) < \pi'(e_t)$. 
But since they are parallel, they have the same endpoints, 
therefore, by the definition 
of $\VO^{\pi'}$ and of $\MO^{\pi'}$, the vertex/edge from which the 
call $\MO^{\pi'}(e_{t})$ was made, would have called $\MO^{\pi'}(e_{t-1})$.
Furthermore, with the exception of $\pi'_{k+1}$ in Case (b),
the the only edge along the path $P$ 
that might be a self-loop is $e$. Otherwise, from Observation~\ref{selfLoopReturnTrue.obs}, 
the self-loop will return true, and thus path $P$ will not visit $e$.

We can write $P$ as $P = (\pi^1_{\sigma(\ell)}, \ldots \pi^1_{\sigma(1)})$
where $\sigma(i) = \pi^1(e_i)$, so that $\sigma(\ell) = k+1$
and $\sigma(\ell-1) = k$.
We next define a mapping $\varphi$ between rankings, such that
$\varphi(\pi^1)$ = $\pi^0$, where we shall show that $\pi^0 \in \Pi^{e,0}$, and
that $\varphi$ is one-to-one. The ranking $\pi^0$ is defined as
follows by ``rotating'' the ranks of the edges on $P$ (and leaving
the ranks of all other edges as in $\pi^1$).
Namely, $\pi^0(e_2) = k+1$, $\pi^0(e_1) = k$, and $\pi^0(e_j) =
\sigma(j-2)$ for every $3 \leq j \leq \ell$. For an illustration,
see Table~\ref{tab:ranksTable}.
\begin{table}[h]
\centering
\begin{tabular}{|l| l| l| l| l| l| l|}
\hline
 & $e_\ell$ & $e_{\ell-1}$ & $\ldots$  & $e_3$ & $e_2$ & $e_1=e$\\
\hline
Rank in $\pi^1$ & $\sigma(\ell)=k+1$ & $\sigma(\ell-1)=k$ & $\ldots$ & $\sigma(3)$ & $\sigma(2)$ & $\sigma(1)$ \\
\hline Rank in $\pi^0$ & $\sigma(\ell-2)$ & $\sigma(\ell-3)$ &
$\ldots$ & $\sigma(1)$
& $\sigma(\ell)=k+1$ & $\sigma(\ell-1)=k$ \\
\hline
\end{tabular}
\caption{Ranking of $P = (e_\ell, \ldots , e_1)$ in $\pi^1$ and in $\pi^0$} \label{tab:ranksTable}
\end{table}
We first verify that $\varphi$ is a projection from $\Pi^{\neg e}$ to $\Pi^{e,0}$.
Namely, we need to show that:
\begin{itemize}
    \item  $\pi^0 \in \Pi_k$, i.e., $\pi^0_{k+1}$ and $\pi^0_k$
share an endpoint $v_2(\pi^0)$, and $e = (v_2(\pi^0),v_3(\pi^0))$.
    \item $X^{\pi^0}(v_1(\pi^0),e) = 0$ (that is, the execution
of $\VO^{{\pi^0}}(v_1(\pi^0))$ does not create a call to
$\MO^{{\pi^0}}(e)$). In other words, (the execution of)
$\VO^{{\pi^0}}(v_1(\pi^0))$ does not  visit $e$.
\end{itemize}
The first item directly follows from the definition of $\pi^0$.
We thus turn to the second item. Recall that by our notational
convention, $v_1(\pi^0) = v_d(\pi^0_{k+1},\pi^0_k) = v_d(e_2,e_1)$
(i.e, it is the endpoint that $e_2$ does not share with $e_1$) so
that it is the common endpoint of $e_2$ and $e_3$, i.e.,
$v_c(e_2,e_3)$. Since 
\BEQ 
\pi^0(e_3) = \sigma(1) < \sigma(\ell) = k+1 = \pi^0(e_2)\;, 
\EEQ 
the execution of $\VO^{{\pi^0}}(v_1(\pi^0))$ will visit $e_3$
before visiting $e_2$.  
Since $\pi^0(e) =  k$, during the execution
of $\VO^{{\pi^0}}(v_1(\pi^0))$, the call to $\MO^{{\pi^0}}(e_3)$
will not cause a recursive call to $\MO^{{\pi^0}}(e)$.

Observe that in the execution of $\VO^{{\pi^1}}(v_1(\pi^1))$, the call
to  $\MO^{\pi^1}(e_3)$ creates a recursive call on
$e_2$ (since $e_2$ follows $e_3$ on the path $P$). Therefore, it
must be the case that $\MO^{{\pi^1}}(e')$={\sc false} for every
$e'$ that has a common endpoint with $e_3$ and such that $\pi^1(e') < \sigma(2)$. 
By the definition of $\varphi$, all edges that are not
on the path $P$ have the same ranks in $\pi^0$ and in $\pi^1$.
Therefore, all edges with rank lower than $\sigma(1)$ have the same
rank in $\pi^1$ and in $\pi^0$. It follows that  for every  $e'$
that has a common endpoint with $e_3$ and such that $\pi^1(e') < \sigma(2)$, 
$\MO^{{\pi^0}}(e') = \mbox{\sc false}$. We can conclude that
$\MO^{{\pi^0}}(e_3) = \mbox{\sc true}$ and so $\VO^{{\pi^0}}(v_1(\pi^0))$
returns {\sc true} without visiting $e_1=e$, as required.

\medskip
It remains to show that $\varphi$ is an injection from $\Pi^{\neg e}$ to $\Pi^{e,0}$.
Assume, contrary to the claim, that $\varphi$ is not an injection.
That is, there are two different rankings $\pi^1 \neq \pi^2 \in \Pi^{\neg e}$ 
where $\varphi(\pi^1) = \varphi(\pi^2)$. Let $P^1 =
(e^1_{\ell_1}, e^1_{\ell_1-1} \ldots e^1_{1})$ and $P^2 =
(e^2_{\ell_2}, e^2_{\ell_2-1}\ldots e^2_{1})$ be the paths that
correspond to the sequence of recursive calls to the maximal
matching oracle, in the executions of
$\VO^{\pi^1}(v_1(\pi^1))$ and  $\VO^{\pi^2}(v_1(\pi^2))$
respectively, where $e^1_{1} = e^2_1 = e$, $\pi^1(e^1_{\ell_1})
= \pi^2(e^2_{\ell_2})= {k+1}$ and  $\pi^1(e^1_{\ell_1-1}) =
\pi^2(e^2_{\ell_2-1})= k$ 
(recall that if $\pi^1$ corresponds to Case (b), then $e^1_{\ell_1}$ is a self-loop
and is not actually part of the sequence of recursive calls that reaches $e$,
and an analogous statement holds for $\pi^2$).
Let $s$
 be the largest index such that
$(e^1_{s}, e^1_{s-1} \ldots e^1_{1}) = (e^2_{s}, e^2_{s-1}\ldots
e^2_{1})$. We denote this common subsequence by $(e_{s}, e_{s-1}
\ldots e_{1})$. Observe that $s \geq 2$. This is true since: (1) By
the definitions of the paths, $e^1_1 = e^2_1 = e$, and (2) given
that $\varphi(\pi^1) = \varphi(\pi^2) = \pi^0$ and
$\pi^0(e^1_2) = \pi^0(e^2_2) = k+1$, it holds that $e^1_2 = e^2_2$.

By the definitions of $\varphi$ and $s$ we have that $\pi^1(e_i)
= \pi^2(e_i)$ for each $i\in [s-2]$. Thus, $\sigma_1(i) =
\sigma_2(i)$ for each $i\in [s-2]$, where we shall sometimes use the
shorthand $\sigma(i)$ for this common value. For an illustration,
see Table~\ref{tab:ranksTableAB}.
\begin{table}[h]
\centering
\begin{tabular}{|l| l| l|}
\hline
 & Rank from $\varphi(\pi^1)$ & Rank from $\varphi(\pi^2)$\\
\hline
 $\pi^0(e_1)$ & $\sigma_1(\ell_1 - 1) = k$ & $\sigma_2(\ell_2 - 1)=k$\\
 \hline
  $\pi^0(e_2)$ & $\sigma_1(\ell_1)=k+1$ & $\sigma_2(\ell_2)=k+1$\\
 \hline
 $\pi^0(e_3)$ & $\sigma_1(1)$ & $\sigma_2(1)$\\
 \hline
 $\vdots$ & $\vdots$ & $\vdots$ \\
 \hline
$\pi^0(e_{s-1})$ & $\sigma_1(s-3)$ & $\sigma_2(s-3)$\\
 \hline
$\pi^0(e_{s})$ & $\sigma_1(s-2)$ & $\sigma_2(s-2)$\\
\hline
\end{tabular}
\caption{\small Ranks of edges $e^1_1 = e^2_1 \ldots e^1_{s-2} =
e^2_{s-2}$ are equal in $\pi^1$ and $\pi^2$}.
 \label{tab:ranksTableAB}
\end{table}

The next observation will be useful.
\begin{obser} \label{lowerRanks.obs}
For every edge $e'$, if $\pi^1(e') < \min\{\sigma_1(s-1),
\sigma_2(s-1)\}$ or $\pi^2(e') < \min\{\sigma_1(s-1),
\sigma_2(s-1)\}$, then $\pi^1(e') = \pi^2(e')$. Therefore,
$\MO^{\pi^1}(e') = \MO^{\pi^2}(e')$ for $e'$ such that
$\pi^1(e')= \pi^2(e') < \min\{\sigma_1(s-1),
\sigma_2(s-1)\}$.
\end{obser}
We consider two cases:
\begin{enumerate}
\item $P^2$ is a suffix of $P^1$ or $P^1$ is a suffix of $P^2$.
Without loss of generality, assume that $P^2$ is a suffix of $P^1$,
so that $s = \ell_2$.
\item Otherwise (neither path is a suffix of the other),
assume without loss of generality that $\sigma_1(s-1) <
\sigma_2(s-1)$.
\end{enumerate}
In both cases, since $e^1_{s+1}$ is not on $P^2$, $\varphi$, when
applied to $\pi^2$ does not change the ranking of $e^1_{s+1}$.
That is, $\pi^0(e^1_{s+1}) = \pi^2(e^1_{s+1})$.
Since (by the definition of $\varphi$)
 $\pi^0(e^1_{s+1}) = \sigma_1(s-1)$,
we get that 
\BEQ\label{pi1b-ea.eq} 
\pi^2(e^1_{s+1}) =
\sigma_1(s-1) = \pi^1(e^1_{s-1})\;. 
\EEQ 
In the first case
(where $P^2$ is a suffix of $P^1$), we have that 
$\sigma_2(s-1) = k$, while $\sigma_1(s-1)< k$,
and so 
\BEQ 
\sigma_1(s-1) < \sigma_2(s-1) \;(=\pi^2(e^2_{s-1}))\;. 
\EEQ
 In the second case, this inequality
was made as an explicit assumption.

 \begin{figure*}[htb]
\begin{center}
\includegraphics{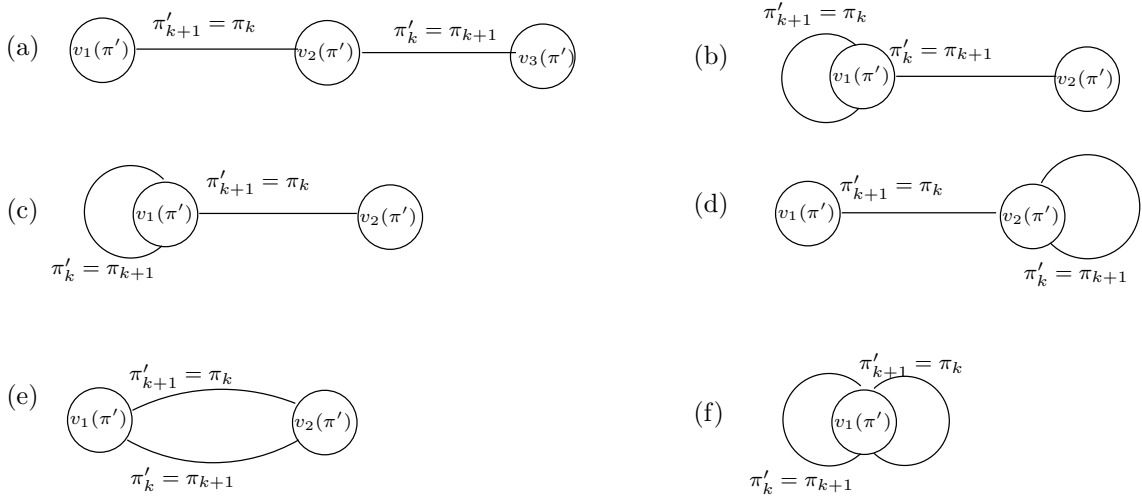}
\end{center}
\caption{\small An illustration for the proof of
Claim~\ref{inject.clm}.}
\end{figure*}

We thus have that the execution of $\MO^{\pi^2}(e^2_s)$ visits
$e^1_{s+1}$ before visiting $e^2_{s-1}$. We would like to understand 
what occurs in the call to $\MO^{\pi^2}(e^1_{s+1})$. If we are in Case (b) 
and $P^1 = (\pi_{k+1}^1, \pi_{k}^1, e)$, i.e., $s = k$,
then, since $e^1_{s+1} = \pi_{k+1}^1$ is a self-loop, from 
Observation~\ref{selfLoopReturnTrue.obs}, $\MO^{\pi^2}(e^1_{s+1}) = \mbox{\sc true}$. 
Hence $\MO^{\pi^2}(e^2_s=e^1_s)$ returns {\sc false} without visiting 
$e^2_{s-1} = e$, but this stands in contradiction to the definition of
$P^2$.
If we are in Case (a), then since the path $P^1$
corresponds to a sequence of recursive calls to the maximal-matching
oracle, we have that for every edge $e'$ that shares an end-point
with $e^1_{s+1}$ and such that
$\pi^1(e') < \sigma_1(s) = \pi^1(e^1_s)$, the call to
$\MO^{\pi^1}(e')$ returns {\sc false}. Combining this with
Observation~\ref{lowerRanks.obs}, we get that for
every edge  $e'$ that shares an end-point with $e^1_{s+1}$
 and such that $\pi^2(e') < \sigma_1(s)$, the call
to $\MO^{\pi^2}(e')$ returns {\sc false}. By
Equation~(\ref{pi1b-ea.eq}) we get that
$\MO^{\pi^2}(e^1_{s+1})$ returns {\sc true}. Hence,
$\MO^{\pi^2}(e^2_s=e^1_s)$ returns {\sc false} without visiting
$e^2_{s-1}$, but this stands in contradiction to the definition of
$P^2$.
\EPF~(Claim~\ref{inject.clm})

\noindent Having established Claim~\ref{inject.clm}, the proof of
Lemma~\ref{Xk-def.lem} is completed. \EPFOF

\bigskip
We are now ready to prove Theorem~\ref{avg-calls.thm}.

\BPFOF{Theorem~\ref{avg-calls.thm}}
Recall that  $d$ denotes the maximum degree,
$\bd$ denotes the average degree and $\rho$ denotes
the ratio between the maximum degree
and the minimum degree, which is denoted by  $d_{\min}$  (where the
latter is at least $1$ since we assumed without loss of generality that
there are no isolated vertices).
By combining
Equations~(\ref{Npiv-Xpiv.eq}) and~(\ref{sumXk.eq}) and applying
Lemma~\ref{Xk.lem} (as well as recalling that we counted each
self-loop as contributing $2$ to the degree of a vertex), we get that:
\BEQN \lefteqn{\frac{1}{m!}\cdot
\frac{1}{n} \cdot \sum_{\pi\in \Pi}\sum_{v\in V} N(\pi,v) }
                    \nonumber\\
&\leq& \frac{1}{m!}\cdot \frac{1}{n} \cdot \frac{1}{2d_{\rm min}} \sum_{e\in E}\sum_{k=1}^m X_k(e)\\
&\leq& \frac{1}{m!}\cdot \frac{1}{n} \cdot \frac{1}{2d_{\min}} \cdot m
 \cdot \left(m\cdot 2(m-1)! + \frac{m\cdot m-1}{2}\cdot (m-2)!\cdot d\right) \\
&=& O\left(\frac{m}{n}\cdot \frac{d}{d_{\min}}\right) \;=\; O(\rho\cdot \bd)\;,
\label{O(d)RecursiveCalls} \EEQN
and we obtain the bound claimed.
\EPFOF

%% file: mm-short.tex
\section{Limiting the Exploration of Neighbor Sets}
\label{mm.sec}

The analysis in the previous section suffices to show an algorithm 
whose query complexity and running time
are a factor of $d$ larger than the number of oracle calls that it makes.
The factor of $d$ in this expression is due to querying all edges that are
incident to the endpoints of each edge for which a call to the maximal
matching oracle is made (where, as we explain 
momentarily,
a random ranking can be selected in an online fashion).

This section is devoted to a method for
selecting incident edges of low rank efficiently without querying 
entire neighborhoods of relevant vertices. By applying the method, 
we reduce the query complexity and the running time by a factor of almost $d$. 
The main challenges here are to ensure that the ranks of encountered edges 
indeed come from the uniform distribution over all permutations and that the same 
decision with respect to a rank of an edge is made at both endpoints of the edges.
 
\ifnum\conf=1
\vspace{-1.5ex}
\fi
\paragraph{Replacing a random ranking with random numbers.}
The oracle construction described as Algorithm~\ref{O_V.alg} and Algorithm~\ref{alg:O_M}
uses a random ranking $\pi:E \to [m]$ of edges. 
We start by replacing a random
ranking of edges with random real numbers in $(0,1]$ selected uniformly and 
independently for every edge $e \in E$,
yielding a vector $\sigma:E \to (0,1]$ which
we use in the same way as the ranking $\pi$. 
Since the probability 
that two edges are assigned the same real number is $0$,
whenever the oracle compares the ranks 
of two edges $e$ and $e'$, 
it can check whether $\sigma(e) < \sigma(e')$,
instead of whether $\pi(e) < \pi(e')$,
effectively yielding a random ranking of edges. 
Since each $\sigma(e)$ is independent,
this small conceptual shift allows one to
generate $\sigma(e)$ at random in an easier manner
and to simplify the analysis. 
Though it is not possible to generate and store 
real numbers in $(0,1]$, we later introduce a proper discretization.

\ifnum\conf=1
\vspace{-1.5ex}
\fi
\subsection{A Data Structure for Accessing Neighbors}
The oracle described as Algorithms~\ref{O_V.alg} and~\ref{alg:O_M} always collects all edges
around the vertex or edge being considered and sorts them to explore them recursively in 
increasing order of their random numbers. In this section we introduce 
a data structure that is responsible 
for generating the random numbers and providing edges for 
further exploration in the desired order.

For every vertex $v \in V$, we have a copy ${\ds[v]}$ of the data structure. 
(In fact, a copy for a given vertex is created when it is accessed for the very first time.) 
From the point of view of the exploration algorithm, the data structure 
{\em exposes only one operation\/}: 
$\lowest(k)$, where $k$ is a positive integer. The
operation ${\ds[v]}.\lowest(k)$ lists edges incident to $v$ in order of 
the random numbers assigned to them, omitting all appearances of parallel 
edges or self-loops except the first one, which has been assigned the 
lowest number. For each positive $k$, the operation 
returns a pair $\langle w,r \rangle$, where $w$ is a vertex and $r$ is 
a number in $(0,1]\cup \{\infty\}$.  If $r\ne\infty$, then $(v,w)$ is 
the edge with the $k^{\rm th}$ lowest number in the above order, 
and $r$ is the number assigned to it. Otherwise, the list is shorter than $k$ 
and $r = \infty$ indicates the query concerned a non-existing edge.
\ifnum\conf=0
 We present the implementation of the data structure in Section~\ref{sec:Data Structures}.
\fi

We rewrite Algorithms~\ref{O_V.alg} and~\ref{alg:O_M} to use the data structure,
and present them as the oracle $\VO^{\sigma}(v)$ in Algorithm~\ref{O_V.alg.2}
and the oracle $\MO^{\sigma}(e)$ in Algorithm~\ref{alg:O_M.2}, respectively.

\begin{algorithm}[ht]
i:=1\;
$\langle w, r \rangle$ := ${\ds[v]}.\lowest(i)$\;
\While {$r \ne \infty$}{
\ifnum\conf=0
     \If { $\MO^{\sigma}((v,w)) =$~{\sc true}}{ \Return {\sc true}}
\else
  {\bf if} $\MO^{\sigma}((v,w)) =$~{\sc true} {\bf then return} {\sc true}\;
\fi
    $i := i+1$\;
    $\langle w, r \rangle$ := ${\ds[v]}.\lowest(i)$\;
}
\Return {\sc false}

\caption{\small An oracle \boldmath$\VO^{\sigma}(v)$ \small for a vertex
cover based on the input from the data structures ${\ds}$, which assigns edges
$e$ random numbers $\sigma(e)$ (online).
Given a vertex $v$, the oracle returns {\sc true} if $v$ belongs to the corresponding
vertex cover and it returns {\sc false} otherwise.}
\label{O_V.alg.2} 
\end{algorithm}

\begin{algorithm}[ht]
\ifnum\conf=0
\If {$\MO^\sigma((u,v))$ has already been computed}{\Return the computed answer}
\else
{\bf if}  $\MO^\sigma((u,v))$ has already been computed {\bf then return} the computed answer
\fi

$k_1$ := $1$ and $k_2$ := $1$ 

$\langle w_1,r_1\rangle$ := \ds[u].\lowest($k_1$) 
             \label{firstW1}

$\langle w_2,r_2\rangle$ := {\ds[v]}.\lowest($k_2$) 
   \label{firstW2}

\While{\rm $w_1 \ne v$ or $w_2 \ne u$}{\label{while} \If{$r_1 <
r_2$}{\label{if_rs}

\lIf{$\MO^{\sigma}((u,w_1))= $ {\sc true}} {\Return{\sc false}} 

$k_1$ := $k_1 + 1$ 

\label{increaseK1} $\langle w_1,r_1\rangle$ :=\ds[u].\lowest($k_1$) 
   \label{newW1}

} \Else{\label{else_rs}

\lIf{$\MO^{\sigma}((v,w_2))=$ {\sc true}} {\Return{\sc false}} 

$k_2$ := $k_2 + 1$\; $\langle w_2,r_2\rangle$ := \ds[v].\lowest($k_2$) 

} }

\Return {\sc true}

\caption{\small An oracle \boldmath$\MO^{\sigma}((u,v))$ \small for a maximal
matching based on the input from the data structures ${\ds}$, which 
assigns edges $e$ random numbers $\sigma(e)$ (online).
Given an edge $(u,v)$, the oracle returns {\sc true} if $(u,v)$ belongs to the corresponding matching
and it returns {\sc false}, otherwise.}
\label{alg:O_M.2}
\end{algorithm}

\noindent
\ifnum\conf=1
The next claim follows from the definition of  {\lowest} (and its proof can
be found in 
\ifnum\hideappendix=1
the full paper).
\else
Appendix~\ref{mm.app}).
\fi
\fi
\begin{claim}\label{Alg-eq.clm}
Let $\sigma$ be an injection from $E$ to $(0,1]$. Let $\pi : E \to [|E|]$
be the corresponding ranking defined in such a way
that for every edge $e$, $\sigma(e)$ is the $\pi(e)^{\rm th}$ lowest number 
in the set $\{\sigma(e'):e' \in E\}$.

For every vertex $v$, the answer returned by 
$\VO^{\sigma}(v)$ (Algorithm~\ref{O_V.alg.2})
is the same as the answer returned by 
$\VO^{\pi}(v)$ (Algorithm~\ref{O_V.alg})
provided 
the operation $\lowest$ works
as specified and gives answers consistent with $\sigma$.
\end{claim}

\def\AlgEqPrf{
It is easy to verify that the claim holds when there are no parallel edges.
This is true because when there are no parallel edges, a sequence of calls to
$\ds[v].\lowest(1),\ldots,\ds[v].\lowest(k)$ simply returns the first $k$ edges
incident to $v$ in order of increasing rank. Furthermore, 
when called on an edge $(u,v)$, Algorithm~\ref{alg:O_M.2} effectively merges
the two corresponding lists of adjacent edges (i.e., those incident to
$u$ and those incident to $v$) to obtain a single list sorted
according to rank, and makes recursive calls in the order dictated by the list.

It remains to verify that the same is true when there are parallel edges. 
For a fixed choice of $\sigma$ and the induced ranking $\pi$
consider the two trees of recursive calls when 
calling $\VO^{\pi}(v)$ (Algorithm~\ref{O_V.alg}) and
$\VO^{\sigma}(v)$ (Algorithm~\ref{O_V.alg.2}), where the former
calls the oracle $\MM^\pi$ (Algorithm~\ref{alg:O_M}), and the latter
calls the oracle $\MM^\sigma$ (Algorithm~\ref{alg:O_M.2}).
When we refer to an edge in in the tree we actually mean 
an occurence of an edge in $G$ on a path of recursive calls.

These trees are both rooted at $v$, and with each edge there is an
associated rank (number) and an associated answer computed by
the corresponding  maximal matching oracle. Recall that each path 
of recursive calls from
the root to a leaf passes through edges with decreasing ranks (numbers).
Furthermore, in both trees,
if an edge $(u,v)$ in the tree is associated with the answer {\sc false}, 
then there must be an edge $(u,w)$ (or $(v,w)$) adjacent to it in the tree with lower rank 
(a ``child'' of this edge) that is associated with the answer {\sc true}, and
it is the highest ranking child that  $(u,v)$  has. If $(u,v)$ is associated
 with the answer {\sc true}, then all the children of
 $(u,v)$ in the tree are associated with the answer {\sc false}.
It will actually be convenient to consider the full recursion trees without
the ``memoization'' rule that we employ (which says that once an answer is determined
for an edge it is not computed again). This in particular implies that 
for each edge that is the last edge on a path of recursive calls, the
answer associated with it must be {\sc true}.

By the definition of  $\VO^{\sigma}(v)$ and the operation $\lowest$,
the tree corresponding to $\VO^{\sigma}(v)$ contains only edges that
have minimal ranking among each set of parallel edges that connect a pair
of vertices. We claim that this tree is a ``pruned'' version of the
tree that corresponds to $\VO^{\pi}(v)$, in the sense that all subtrees
containing non-minimally ranked parallel edges are removed, and the
answers associated with the remaining edges (and hence with the
root $v$) are exactly the same.

Let $T^\pi(v)$ denote the tree of recursive calls for $\VO^{\pi}(v)$,
and let $h$ be the height of $T^\pi(v)$. Starting from $\ell=h$ and going
up the tree, we show that we can remove all non-minimally ranked
parallel edges in level $\ell$ of $T^\pi(v)$
without altering the answer for their parent edges.
For $\ell=h$, we claim that there are no non-minimally ranked
parallel edges in the last level of $T^\pi(v)$,
so that no pruning needs to be performed.
To verify this, assume in contradiction that 
$e$ is a non-minimally ranked parallel edge between vertices $u$ and
$w$ where $e$ is at the end of a recursive path of length $h$ in $T^\pi(v)$.
Since $e$ is not minimally ranked, 
there should be a ``sibling'' of $e$ in the tree
which correspond to the minimally ranked edge  $e'$ between $u$ and $w$.
 Since $\pi(e') < \pi(e)$, it must be the case that
 the answer associated with $e'$, that is, $\MM^\pi(e')$, is {\sc false}.
But $e'$ also belongs to level $h$, so that $e'$ is the last edge on a path of
recursive calls, and hence cannot be answered  {\sc false}.

Assuming we have performed the pruning successfully for all levels
$\ell < \ell' \leq h$, we show that we can perform it for level $\ell$.
Consider a non-minimally ranked parallel edge $e$ between vertices $u$
and $v$ in level $\ell$ of $T^\pi(v)$. As argued above, 
there is a ``sibling'' of $e$ in the tree
which correspond to the minimally ranked edge  $e'$ between $u$ and $w$. 
 Since $\pi(e') < \pi(e)$, it must be the case that
 the answer associated with $e'$, that is, $\MM^\pi(e')$, is {\sc false}.
This implies that $e'$ has a child $e''$ in the tree resulting from
pruning all non-minimal parallel edges from levels $\ell' > \ell$,
such that $\MM^\pi(e'') =  \mbox{\sc true}$. But since $\pi(e'') < \pi(e') < \pi(e)$,
and $e''$ is also adjacent to $e$, we get that $\MM^\pi(e)$ is {\sc false}
as well. Hence, it is possible to prune $e$ from the tree without altering
the answer obtained for its parent. 
}
\ifnum\conf=0
\BPF
\AlgEqPrf
\EPF
\fi

\ifnum\conf=1
For the sake of simplicity of the presentation, from this point on
we consider the case in which there are no parallel edges. We show how 
to deal with parallel edges 
\ifnum\hideappendix=1
in the full version of the paper.
\else
in Appendix~\ref{mm.app}.
\fi
\fi

\ifnum\conf=1
\vspace{-1.5ex}
\subsection{Implementing {\lowest}}\label{lowest.subsec}
\else
\subsection{Implementing {\lowest}: The High-Level Idea}\label{lowest.subsec}
\fi
The  pseudo-code for the procedure {\lowest} as well as the data structure
that it uses, are given in full detail in 
\ifnum\conf=1
\ifnum\hideappendix=1
the full paper.
\else
Appendix~\ref{sec:Data Structures}.
\fi
\else
Subsection~\ref{sec:Data Structures}.
\fi
Here we give a high-level description.
\ifnum\conf=0
For the sake of simplicity of the presentation, 
in this description we assume that there are no parallel edges.
\fi

Roughly speaking, the procedure {\lowest} for a vertex $v$
is implemented in ``batches''. Namely, considering intervals of $(0,1]$ 
of the form $(2^{-i},2^{-i+1}]$ (for $i\in[d_\star]$,
where $d_\star = \left\lceil \log d \right\rceil$, as well as the interval
$(0,2^{-d_\star}]$), the procedure does the following. It first
decides which edges incident to $v$ 
should be assigned a value in the current interval
$(2^{-i},2^{-i+1}]$. In this stage each edge is identified
with its label (in $\{1,\ldots,\deg(v)\}$).
The procedure then determines the identity of
the other endpoint of each of these edges by performing a neighbor query,
and it assigns the edge a value $\sigma((v,w))$, selected uniformly at random from the
interval. This assignment is performed unless a certain constraint is
discovered due to information held in $\ds[w]$, as we explain
subsequently. Once $\sigma((v,w))$ is determined, the other endpoint of
the edge, $w$, is ``notified''. That is, the data structure $\ds[w]$
is updated with this new information.  The procedure ``opens''  a new
interval $(2^{-i+1},2^{-i+2}]$ if the index $k$ it is called with is such that
the number of neighbors $w$ of $v$ whose identity has been revealed
and such that $\sigma((v,w)) \leq 2^{-i+1}$ is strictly less than $k$.
Thus, the procedure performs queries and assigns valued to edges ``on demand'', but
it does so for ``batches'' of edges. 
More precise details follow.

The data structure  {\ds} maintains
two values for each vertex $v$: {\tt lb}, and {\tt next\_lb}
(where the latter is always twice the former).
When a vertex is first encountered, {\tt lb} is set to $0$ and
{\tt next\_lb} is set to $2^{-d_\star}$. 
Second, the data structure maintains a dictionary 
{\tt assigned\_number}, which holds, for those vertices $w$ that
are known to be neighbors of $v$, the value $\sigma((v,w))$ that 
was assigned to the edge between them (initially, the dictionary is empty).
The subset of indices in
 $\{1,\dots,\deg(v)\}$ that correspond to edges for which the other
endpoint has not been revealed (and do not yet have an associated value),
are considered {\em unassigned\/}.
Third, the data structure maintains a list of pairs
 $\langle w,r \rangle$,
where $w$ is a  (known) neighbor of $v$ and $r = \sigma((v,w))$.
This list is sorted in ascending order of $r$'s, and it contains exactly those
$w$ for which the corresponding $r$ is at most {\tt lb}.

If a call is made to ${\ds[v]}.\lowest(k)$ with $k > \deg(v)$ then
it returns\footnote{Recall that we assume that there are no
parallel edges, or else $\langle v,\infty\rangle$ is returned if $k$
exceeds the ``effective'' degree of $v$, that is, counting parallel
edges as a single edge.}
$\langle v,\infty\rangle$.
Otherwise, the
procedure does the following until the length of {\tt sorted} is at least $k$.
It first considers those edges $(v,w)$ incident to $v$ that were already
assigned a value $r$ and this value belongs to the interval
$(\mbox{\tt lb},\mbox{\tt next\_lb}]$ (that is,
$\mbox{\tt assigned\_number}[w] \in (\mbox{\tt lb},\mbox{\tt next\_lb}]$).
The setting of the value $r$ for each such edge $(v,w)$ was performed previously
in the course of call to ${\ds[w]}.\lowest(\cdot)$. 
Let the corresponding subset of pairs $\langle w,r \rangle$ be denoted $S$.

The procedure next selects a subset $T$ of $\{1,\ldots,\deg(v)\}$ containing the
labels of those (additional) edges that it will (tentatively) assign a value in
({\tt lb},{\tt next\_lb}]. Putting aside for now the issue of time-efficiency
(which we return to later), this can be done by flipping a coin with bias 
$\frac{\mbox{\tt next\_lb} - \mbox{\tt lb}}{1 - \mbox{\tt lb}}$
independently for each edge label in the subset of unassigned edge labels.
For each $t \in T$, the procedure now performs a neighbor query to obtain
the $t^{\rm th}$ neighbor of $v$. Denoting this neighbor by $w$, 
let {\tt lb}' denote the lower bound {\tt lb} held by $w$, that is, in the 
data structure $\ds[w]$. If $\mbox{\tt lb}' \leq  {\tt lb}$, so that the
lower bound constraint imposed by $w$ is no larger than that imposed by $v$,
then the following operations are performed.

First, a random number $r$ in the interval ({\tt lb},{\tt next\_lb}]
is selected uniformly at random, and $\mbox{\tt assigned\_number}[w]$ is set
to $r$. In addition, $\mbox{\tt assigned\_number}[v]$ is set to $r$
in the data structure ${\ds[w]}$ (so that $w$ is ``notified'' of the revealed edge
$(v,w)$ as well as the assignment $r = \sigma((v,w))$).  
Finally, the pair $\langle w , r \rangle$ is added to $S$. 

If $\mbox{\tt lb}' > \mbox{\tt lb}$, which means that $\mbox{\tt lb}'\geq \mbox{\tt next\_lb}$
(given the way the intervals are defined), then the lower bound constraint
imposed by the end point $w$ of the edge $(v,w)$ does not allow the edge
to be assigned a value in the interval ({\tt lb},{\tt next\_lb}], and so 
effectively its selection to $T$ is retracted. Note that since the decision
whether an edge label is added to $T$ is done independently for the different edges,
the end effect (of not assigning $(v,w)$ a value in
 ({\tt lb},{\tt next\_lb}]) is exactly the same as the one we would get 
if we had the knowledge in advance (before selecting $T$), that the corresponding
edge label $t$ should not be selected.

After going over all labels $t$ in $T$, the resulting set $S$ of pairs $\langle w , r \rangle$
is sorted in ascending order of $r$'s, and it is appended to the end of the list
{\tt sorted}. Thus, {\tt sorted} now includes all pairs  $\langle w , r \rangle$ such that
$w$ is a neighbor of $v$, the value assigned to this edge is $r$, and $r \leq \mbox{\tt next\_lb}$.
The variables {\tt lb} and {\tt next\_lb} are then updated so that
{\tt lb} is set to {\tt next\_lb} and {\tt next\_lb} is set to $2\cdot  \mbox{\tt next\_lb}$.
Once the length of {\tt sorted} is at least $k$, the procedure returns
$\mbox{\tt sorted}[k]$.
\ifnum\conf=1
\ifnum\hideappendix=1
In the full paper 
\else
In Appendix~\ref{sec:Data Structures}
\fi
\else
In Subsection~\ref{sec:Data Structures}
\fi
we formally establish
that the distribution of random numbers the data structures $\ds[v]$ provide access to
via the operation $\lowest(k)$ is the same as assigning independently at random
a number from the range $(0,1]$ to each edge.

\ifnum\conf=1
\vspace{-1.5ex}
\subsection{Query Complexity}
We show that the number of queries that the algorithm makes is not 
much higher than the number of recursive calls in the graph exploration procedures. 
More precisely, if the expected number of calls to the oracles
(Algorithm~\ref{O_V.alg} and Algorithm~\ref{alg:O_M}) is upper bounded by $t$,
then with high constant probability (over the choice of the random assignment of
values to edges) the number of queries performed (when using the oracles
as in Algorithm~\ref{O_V.alg.2} and Algorithm~\ref{alg:O_M.2} and implementing $\lowest$
as described above) is $O(t \cdot \log^2 (dt))$. The full analysis is given in
\ifnum\conf=1
\ifnum\hideappendix=1
the full paper, and here we give the high level idea.
\else
Appendix~\ref{query.app}, and here we give the high level idea.
\fi
\else
Subsection~\ref{query.app}, and here we give the high level idea.
\fi

Consider an idealized scenario in which, for all vertices $v$,
the number of edges incident to $v$
that are assigned a number in ({\tt lb},{\tt next\_lb})
(for all intervals ({\tt lb},{\tt next\_lb}) such that a call is made to {\tt lowest} when
({\tt lb},{\tt next\_lb}) is the current interval), is {\em exactly\/} the 
expected value. Furthermore, assume that in the process of assigning
random numbers to edges, no conflict is detected between 
endpoints of any edge (in terms of the lower bounds on the random number
that should be assigned to the edge).
 In such a case, since each interval is double the size
of the previous interval, the number of queries performed is
at most twice the number of calls to the maximal matching oracle.
In reality, there can of course be deviations from this expected
value, and conflicts may occur.  However, we can upper bound
the probability of large deviations (where we need to take into
account deviations in both directions), and thus obtain the 
upper bound on the query complexity claimed above.

\ifnum\conf=1
\vspace{-1.5ex}
\fi
\subsection{Running Time}
Given the bound on the query complexity of the algorithm, we get a
bound that is not much larger on the number of operations which can be viewed
as ``accompanying'' each query. By this we mean: selecting a random
number for the edge, updating the data structures of the endpoints of the
edge with this number, and sorting the list {\tt sorted}.
However, we also need to implement the random choice of edge labels that
are put in the set $T$ (of (tentative) labels of edges that are supposed
to be assigned random numbers in the current interval ({\tt lb},{\tt next\_lb}]).
When we ignored the issue of running time (in Subsection~\ref{lowest.subsec})
we explained how this can be done in time $O(d)$. However, a more sophisticated
implementation gives us the next theorem.

\begin{theorem}\label{thm:running_time}
Consider an algorithm $\mathcal A$ that queries the input graph only via the oracle described as Algorithm~\ref{O_V.fig}. Let $t \ge 1$ be a bound on the expected resulting number of calls in $\mathcal A$ to the oracles described as Algorithm~\ref{O_V.fig} and Algorithm~\ref{alg:O_M}, and such that $t$ fits into a constant number of machine words using the standard binary representation
\footnote{Throughout the paper we silently assume the \emph{Word RAM model}, which attempts to model the real-world computation. In this model, every memory cell (we call them \emph{machine words}) is a $w$-bit integer and standard arithmetic operations on memory cells are performed in $O(1)$ time. Moreover, the number of bits in the machine word
is large enough to address any part of the input. In our case, this implies that $w = \Omega(\log n)$.

The issues related to the size of the machine word occur in this paper when instead of real numbers we keep their approximate values.}. Let $d$ be an upper bound on the maximum degree of the input graph.

Suppose that calls to Algorithm~\ref{O_V.fig} are replaced with calls to Algorithm~\ref{O_V.alg.2}.
The oracles described as Algorithm~\ref{O_V.alg.2} and Algorithm~\ref{alg:O_M.2} can be implemented in such a way that with probability $4/5$ all of the following events hold:
\begin{itemize}
\item The number of queries to the graph is $O(t \cdot \log^2 (dt))$.
\item The total time necessary to compute the answers for the queries to the oracles is $O(t \cdot \log^3 (dt))$.
\item The distribution of the answers that the oracle gives is $\mathcal D$ such that
for some other distribution $\mathcal D'$ over answers, the convex combination $\frac{4}{5}\cdot\mathcal D + \frac{1}{5}\cdot \mathcal D'$ is the distribution of answers of the oracle described as Algorithm~\ref{O_V.fig}.
\end{itemize}
\end{theorem}
\fi

%% file: mm.tex

\ifnum\conf=1

\noindent{\bf Claim~\ref{Alg-eq.clm} (Restated)}~
{\em
Let $\sigma$ be an injection from $E$ to $(0,1]$. Let $\pi : E \to [|E|]$
be the corresponding ranking defined in such a way
that for every edge $e$, $\sigma(e)$ is the $\pi(e)^{\rm th}$ lowest number
in the set $\{\sigma(e'):e' \in E\}$.

For every vertex $v$, the answer returned by
$\VO^{\sigma}(v)$ (Algorithm~\ref{O_V.alg.2})
is the same as the answer returned by
$\VO^{\pi}(v)$ (Algorithm~\ref{O_V.alg})
provided
the operation $\lowest$ works
as specified and gives answers consistent with $\sigma$.
}

\medskip

\BPF
\AlgEqPrf
\EPF
\fi

\subsection{Generating Random Numbers}
\label{sec:random_process}

In this subsection we describe a random process that generates random
numbers $\sigma(e)$ for edges $e \in E$.
The procedure $\lowest$ applies this process in the course of its executions.
In the remainder of this section, $|\mathcal I|$
denotes the length of an arbitrary real interval $\mathcal I$.
We do not distinguish open and closed intervals here. For instance,
$|(0,1)| = |[0,1]| = |(0,1]| = |[0,1)| = 1$.

Let $d$ be an upper bound on the maximum vertex degree.
We set $d_\star = \left\lceil \log d \right\rceil$.
For every edge $e$, the number $\sigma(e)$ should be selected
independently, uniformly at random from the range $(0,1]$.
We partition this range into $d_\star+1$ intervals. We set
$$
\mathcal I_i = \begin{cases}
(2^{-i},2^{-i+1}]& \mbox{for $i\in[d_\star]$,}\\
(0,2^{-d_\star}]& \mbox{for $i=d_\star+1$.}
\end{cases}
$$

\begin{algorithm}[ht]
\label{ds:choosingIntervalFirst}
\caption{A Process for Selecting a Random Number Assigned to an Edge}
\SetKwInOut{Input}{input}\SetKwFunction{Break}{Break}
\For {$i\leftarrow d_{*}+1$ \KwDownTo $2$}{
with probability $\frac{|\mathcal I_i|}{\sum_{1 \leq
j\leq i}|\mathcal I_j|}$: \Return a number selected from $\mathcal I_i$ uniformly at random
(and terminate)
}
\Return a number selected from $\mathcal I_1$ uniformly at random
\end{algorithm}

We describe our process as Algorithm~\ref{ds:choosingIntervalFirst}.
The process first selects one of the intervals $\mathcal I_i$, and
then selects a number uniformly at random from this interval.
The selection of the interval is conducted as follows. We consider
the intervals in reverse order, from $\mathcal I_{d_\star+1}$ to $\mathcal I_1$.
For a considered interval, we decide that the number belongs to
this interval with probability equal to the length of the interval
over the sum of lengths of all the remaining intervals. The process
selects each interval with probability proportional to its length,
and since the lengths of all intervals sum up to 1, the number that
the process returns is uniformly distributed on the entire interval $(0,1]$.

\begin{corollary}\label{cor:random_numbers}
Algorithm~\ref{ds:choosingIntervalFirst} selects a random number from
the uniform distribution on $(0,1]$.
\end{corollary}

Note that by simulating a few iterations of the loop in the above process,
one can decide that the number assigned to a given edge is either a
specific number less than or equal to $2^{-i}$, or that it is greater than
$2^{-i}$ without specifying it further, for some $i$. Later, whenever
more information about the number is needed, we may continue with
consecutive iterations of the loop. As we see later, we use the process
in our data structures ${\ds}[v]$ to lower the query and time complexity
of the resulting vertex cover algorithm.

\subsection{Data Structures}
\label{sec:Data Structures}

We now describe the data structures ${\ds[v]}$. Each data structure ${\ds[v]}$
simulates the random process described in Section~\ref{sec:random_process} for
all edges incident to $v$ in the course of the executions of
$\ds[v].\lowest$. The data structure simultaneously makes a single
iteration of the loop in Algorithm~\ref{ds:choosingIntervalFirst} for all
incident edges.
It may be the case that for some edge $(v,w)$, the random number
has already been specified. In this case, the result of the iteration for
this $(v,w)$ is discarded. It may also be the case that this iteration
of the loop has already been taken care of by ${\ds[w]}$, the
data structure for the other endpoint of the edge. The data structures
communicate to make sure that a second execution of a given iteration
does not overrule the first. The data structures are designed to
minimize the amount of necessary communication. Note that if a data structure does not have to communicate with a data structure at the other endpoint of a given neighbor, it does not even have to know the neighbor it is connected to with a given edge, which can be used to save a single query. By using this approach, we eventually save a factor of nearly $d$ in the query complexity.

Each data structure ${\ds[v]}$ supports the following operations:
\begin{description}

\item[\mbox{${\ds[v]}.\lowest(k)$}:] As already mentioned, this is the only
operation that is directly used by the oracles (Algorithm~\ref{O_V.alg.2}
and Algorithm~\ref{alg:O_M.2}).  It returns a pair $\langle w,r\rangle$, where
$(v,w)$ is the edge with the $k^{\rm th}$ lowest random number assigned to
edges incident to $v$, omitting a second and furher appearances for parallel edges,
and $r$ is the random value assigned to $(v,w)$. If $r=\infty$, then $k$
is greater than the length of such a defined list.

\item[\mbox{${\ds[v]}.\getlb()$}:] The operation returns the current lower bound the data structure imposes on the edges that are incident to $v$ and have not been assigned a specific random number yet. The set of possible values returned by the procedure is $\{0\} \cup \{2^{i}:-d_\star \le i \le 0 \}$. Let $\ell_v$ be the number returned by the operation. It implies that the data structure simultaneously simulated the random process described in Section~\ref{sec:random_process} for incident edges until it made sure that the random numbers that have not been fixed belong to $(\ell_v,1]$.

Furthermore, let $(v,w)$ be an edge in the graph. Let $\ell_v$ and $\ell_w$ be the numbers returned by the operation for ${\ds[v]}$ and ${\ds[w]}$, respectively. If no specific random number has been assigned to $(v,w)$, then we know that the random number will eventually be selected uniformly at random from  $(\max\{\ell_v,\ell_w\},1]$.

\item[\mbox{${\ds[v]}.\setvalue(w,r)$}:] It is used to notify the data structure ${\ds[v]}$ that the random value assigned to $(v,w)$ has been set to $r$. This operation is used when the data structure ${\ds[w]}$ assigns a specific random number to $(v,w)$. Before assigning $r$, the data structure ${\ds[w]}$ has to make sure that $r > {\ds[v]}.\getlb()$, i.e., it has not been decided by the data structure ${\ds[v]}$ that the random number assigned to $v$ is greater than $r$.

\end{description}

To implement the above operations, each data structure ${\ds[v]}$ maintains the following information:
\begin{description}
\item[\mbox{\tt lb}:] The variable specifies the lower bound on the incident edges that were not assigned a random number yet. This is the value returned by the operation ${\ds[v]}.\getlb()$. This is also the value at which the simulation of the process generating random number for edges incident to $v$ has stopped.

\item[\mbox{\tt next\_lb}:] If specific random numbers assigned to more edges are necessary, the next considered range of random numbers will be $(\mbox{\tt lb},\mbox{\tt next\_lb}]$, and \mbox{\tt next\_lb} will become the new lower bound for the edges that have not been assigned any random number. This variable is redundant, because its value is implied by the value of \mbox{\tt lb}, but using it simplifies the pseudocode.

\item[\mbox{\tt assigned\_number}:] This is a dictionary that maps neighbors $w$ of $v$ to numbers in $(0,1]$. Initially, the dictionary is empty. If $\mbox{\tt assigned\_number}[w] = \mbox{\sc none}$, i.e., there is no mapping for $w$, then no specific random number has been assigned to any of the edges $(v,w)$ yet. Otherwise,
$\mbox{\tt assigned\_number}[w]$ is the lowest random number that has been assigned to any parallel edge $(v,w)$.

 \item[\mbox{\tt sorted}:] This is a list consisting of pairs $\langle w,r \rangle$,
where $w$ is a neighbor of $v$ and $r$ is the number assigned to the edge $(v,w)$.
It is sorted in ascending order of $r$'s, and it contains exactly those
$w$ for which the edge $(v,w)$ (with the lowest assigned random number)
 has an assigned random number less than or equal to ${\tt lb}$.
For all neighbors $w$ that do not appear on the list, the lowest number
assigned to any edge $(v,w)$ is greater than ${\tt lb}$.

\end{description}

We give pseudocode for all data structure operations as
Algorithms~\ref{ds:init}, \ref{ds:setvalue},
\ref{ds:lower_bound}, and~\ref{ds:lowest}.
We postpone all issues related to an efficient implementation of the data structure to Section~\ref{sec:efficient_implementation}.
Three of them are straightforward, and we only elaborate on the operation ${\ds[v]}.\lowest(k)$ (see
Algorithm~\ref{ds:lowest}).

\begin{algorithm}[th]
\label{ds:init}
\caption{The procedure for initializing ${\ds[v]}$}
{\tt lb} := 0\;
{\tt next\_lb} := $2^{-d_\star}$\;
{\tt assigned\_number} := \{\hbox{empty map}\}\;
{\tt sorted} := \{\hbox{empty list}\}\;
\end{algorithm}

\begin{algorithm}[th]
\label{ds:setvalue} \caption{The procedure ${\ds[v]}.\setvalue(w,r)$}
$\mbox{\tt assigned\_number}[w]$ := $r$
\end{algorithm}

\begin{algorithm}[th]
\label{ds:lower_bound} \caption{The procedure ${\ds[v]}.\getlb()$}
\Return {\tt lb}
\end{algorithm}

As long as not sufficiently many lowest random numbers assigned to edges incident to $v$ have been determined, the operation $\lowest$ simulates the next iteration of the loop in the random process that we use for generating random numbers. Let $I$ be the interval $(\mbox{\tt lb},\mbox{\tt next\_lb}]$. The operation wants to determine all random numbers assigned to edges incident to $v$ that lay in $I$.
First, in Line~2, it determines the random numbers in $I$ that have already been assigned by the other endpoints of corresponding edges. In Line~3, the operation simulates an iteration of the loop of the random process for all edges incident to $v$ to determine a subset of them that will have numbers in $I$ (unless it has already been decided for a given edge that its random number is not in $I$). In the loop in Line~4, the operation considers each of these edges. Let $(v,w)$ be one of them, where $w$ is its other endpoint, queried by the operation. In Line~6, the operation generates a prospective random number $r \in I$ for the edge.
First, the operation makes sure that this iteration of the has not been simulated by the other endpoint (the condition in Step~7). If this is the case, the operation considers two further cases.
If $r$ is lower than the lowest number assigned to any parallel edge $(v,w)$ so far, the procedure updates the appropriate data structures with this information (Steps~8--11). If no random number has ever been assigned to any edge $(v,w)$, the procedure assigns it and updates the data structures appropriately (Step~12--15).
When the operation finishes going over the list of all potentially selected edges and eventually determines all incident edges with new lowest random numbers, it sorts them in order of their random number and appends them in this order to the list $\mbox{\tt sorted}$. Finally, when sufficiently many edges with lowest numbers have been determined, the operation returns the identity of the edge with the $k^{\rm th}$ smallest number.

\begin{algorithm}[ht]
\label{ds:lowest} \caption{The procedure ${\ds[v]}.\lowest(k)$}
\While{\rm $\length(\mbox{\tt sorted}) < k$ and $\mbox{\tt lb} < 1 $}{

  $S$ := \hbox{\rm set of pairs $\langle w,r \rangle$ such that
         $\mbox{\tt assigned\_number}[w] = r$
         and $r \in (\mbox{\tt lb},\mbox{\tt next\_lb}]$}\;

  $T$ := subset of $\{1,\ldots,\deg(v)\}$ with each number included independently\newline
  \phantom{$T$ := }with probability $\frac{\mbox{\tt next\_lb} - \mbox{\tt lb}}{1 - \mbox{\tt lb}}$
  \label{step:generate_subset}\;

  \ForEach{$t \in T$}{

    $w$ := \rm $t^{\rm th}$ neighbor of $v$\;
    $r$ := a number selected uniformly at random from $(\mbox{\tt lb},\mbox{\tt next\_lb}]$\;

    \If{$\ds[w].\getlb() \le \mbox{\tt lb}$}{
	\If{$\exists \langle w,r'\rangle \in S$ s.t. $r < r'$}{
	  $\mbox{\tt assigned\_number}[w]$ := $r$\;
	  $\ds[w].\setvalue(v,r)$\;
	  replace $\langle w,r'\rangle$ with $\langle w,r\rangle$ in $S$
	}
         \If{$\mbox{\tt assigned\_number}[w] = \mbox{\sc none}$}{
         $\mbox{\tt assigned\_number}[w]$ := $r$\;
         $\ds[w].\setvalue(v,r)$\;
         $S$ := $S \cup \{\langle w , r \rangle\}$\;
     }
    }

  }

Sort $S$ in ascending order of their $r$, and append at the end of ${\tt sorted}$\;
\mbox{\tt lb} := \mbox{\tt next\_lb}\;
\mbox{\tt next\_lb} := $2 \cdot \mbox{\tt next\_lb}$\;
}

\lIf{$\mbox{\rm length}(\mbox{\tt sorted}) < k$}{\Return $\langle v,\infty \rangle$}\;
\lElse{\Return $\mbox{\tt sorted}[k]$}\;
\end{algorithm}

\begin{lemma}
The lists of incident edges that the data structures $\ds[v]$ provide access to
via the operation $\lowest(k)$ are distributed in the same way
as when each edge is assigned independently at random a number from the range $(0,1]$.
\end{lemma}

\BPF
We  know from Corollary~\ref{cor:random_numbers} that the random process generates a random number from the distribution $(0,1]$. Each data structure $\ds[v]$ simulates consecutive iterations of the loop in this process for all edges incident to $v$. Consider a group of parallel edges $(v,w)$.
For each of these edges, the random process is simulated by both $\ds[v]$ and $\ds[w]$. We have to show that until the lowest number assigned to the edges in this group is determined (which happens when it is added to the list {\tt sorted}), then for each edge the decision made in the first simulation matters.
Why is this the case?
Recall that the random process considers intervals $\mathcal I_{d_\star+1}$, $\mathcal I_{d_\star}$, \ldots, $\mathcal I_1$ as the sources of the random number in this order.
As long as both $\ds[v]$ and $\ds[w]$ reject a given interval their decisions are the same, so the first decision is in effect. Now suppose without loss of generality that $\ds[w]$ simulates a consecutive iteration of the loop in the random process and decides to use $\mathcal I_{i}$ as the source of the random number for a given edge $(v,w)$ in Step~\ref{step:generate_subset} of the operation $\lowest$. If $\ds[v]$ has already simulated this iteration (the condition verified in Step~7), the operation does not proceed. Otherwise, the random number assigned to the edge is considered for a new minimum random number assigned to this group of parallel edges. Note that
since the operation keeps simulating iterations even after a random number is assigned, it could be the case for a specific copy of $(v,w)$ that a new, higher random number is considered, but it is ignored, because it is higher than the first decision, which is the only one that has impact on the list that the operation $\lowest$ provides access to.

The correctness of the data structure follows from the fact that it extends the list {\tt sorted} by always adding all edges with random numbers in a consecutive interval, and it always takes into consideration decisions already made by data structures for the other endpoints for these intervals.
\EPF

\subsection{Query Complexity}
\label{query.app}

We now show that the number of queries that the algorithm makes is not much higher than the number of recursive calls in the graph exploration procedures. The following simple lemma easily follows from the Chernoff bound and will help us analyze the behavior of the algorithm.

\begin{lemma}\label{lem:expectation_divergence}
Let $X_1$, \ldots, $X_s$ be independent random Bernoulli variables such that each $X_i$ equals $1$ with probability $p$. It holds:
\begin{itemize}
\item For any $\delta \in (0,1/2)$,
 $$\sum_{i}X_i \le 6 \cdot \ln(1/\delta) \cdot \max\{1,ps\}.$$
with probability at least $1-\delta$.
\item
For any $\delta \in (0,1/2)$, if $ps > 8 \ln(1/\delta)$, then
$$\sum_{i}X_i \ge \frac{ps}{2}.$$
with probability at least $1-\delta$.
\end{itemize}
\end{lemma}

\BPF
Let us first prove the first claim. If $6 \cdot \ln(1/\delta) \cdot \max\{1,ps\} \ge s$, the claim follows trivially.
Otherwise, there exist independent Bernoulli random variables $Y_i$, $1 \le i \le s$ such that for each $i$,
\BEQ \Pr[Y_i=1]=3\cdot \ln(1/\delta) \cdot \max\{1/s,p\} > p \nonumber \EEQ
since from the definition of $\delta$: $3 \cdot \ln(1/\delta) > 1$.
Therefore $\Pr[X_i = 1] < \Pr[Y_i = 1]$. By this fact and by the Chernoff bound,
\BEQN \Pr[\sum X_i > 2E[\sum Y_i]] &\le& \Pr[\sum Y_i > 2E[\sum Y_i]] \nonumber\\
&\le& \exp(-\ln(1/\delta) \cdot \max\{1,ps\}) \nonumber \\
&\le& \exp(-\ln(1/\delta)) \le \delta. \nonumber \EEQN
The second claim also directly follows from the Chernoff bound:
 \BEQ \Pr[\sum X_i < ps / 2] \le \exp(-(1/2)^2 \cdot ps / 2) \le \delta.\nonumber\EEQ
\EPF

\begin{definition}
Denote $\mathcal J_i = \bigcup_{j=i}^{d_\star+1} \mathcal I_j$, where $1 \le i \le d_\star+1$.
For example: $\mathcal J_1 = (0,1]$ and $\mathcal J_{d_\star+1} = (0,\frac{1}{d}]$.
We expect that the number of incident edges to $v$ with random numbers in $\mathcal J_i$ to be $\deg(v)\cdot |\mathcal J_i|$.
\end{definition}\label{def:alpha_usual}
We now define a property of vertices that is useful in our analysis. Intuitively, we say that a vertex is ``usual'' if the numbers of incident edges with random numbers in specific subranges of $(0,1]$ are close to their expectations.
\begin{definition}
Let $\alpha > 0$.
We say that a vertex $v$ is $\alpha$-\emph{usual} if the random numbers assigned to edges incident to $v$ have the following properties for all $i \in \{1,\ldots,d_\star+1\}$:
\begin{itemize}
 \item Upper bound: The number of incident edges with random numbers in $\mathcal J_i$
is
\BEQ \mbox{at most } \max\{\alpha,\alpha\cdot\deg(v) \cdot |\mathcal J_i| \}. \nonumber \EEQ
 \item Lower bound: If $\deg(v) \cdot |\mathcal J_i| \ge \alpha$, then the number of edges with random numbers in $\mathcal J_i$ is
\BEQ \mbox{at least } \deg(v) \cdot |\mathcal J_i| / 2. \nonumber \EEQ
\end{itemize}
\end{definition}

We now basically want to show that the relevant vertices are $\alpha$-\emph{usual}, and later on we will use it to prove a lower bound.

We define an additional quantity that is useful later in bounding the total running time of the algorithm.

\begin{definition}
For an execution of Step~\ref{step:generate_subset} of Algorithm~\ref{ds:lowest} where the number of neighbors is $k$ and $p \in [0,1]$ is the probability of selecting each of them, we say that the \emph{toll} for running it is $kp$.
\end{definition}

We now prove a bound on the query complexity of the algorithm and other quantities, which are useful later to bound the running time. We start by introducing the main Lemma (Lemma~\ref{lem:query_complexity}), followed by proving Lemma~\ref{lem:helper_query_complexity} which will help us prove Lemma~\ref{lem:query_complexity}.

\begin{lemma}\label{lem:query_complexity}
Consider an algorithm $\mathcal A$ that queries the input graph only via the oracle described as Algorithm~\ref{O_V.fig}. Let $t \ge 1$ be the expected resulting number of calls in $\mathcal A$ to the oracles described as Algorithm~\ref{O_V.fig} and Algorithm~\ref{alg:O_M}. Let $d$ be an upper bound on the maximum degree of the input graph.

Suppose now that we run this algorithm replacing calls to Algorithm~\ref{O_V.fig}
with calls to Algorithm~\ref{O_V.alg.2}.
The following events hold all at the same time with probability $1-1/20$:
\begin{enumerate}
 \item The total number of calls to Algorithms~\ref{O_V.alg.2} and~\ref{alg:O_M.2} is $O(t)$
 \item The operation $\lowest$ in data structures $\ds[v]$ runs at most $O(t)$ times.
 \item The query complexity of $\mathcal A$ is $O(t \cdot \log^2 (dt))$.
 \item The total toll for running Step~\ref{step:generate_subset} of Algorithm~\ref{ds:lowest} is $O(t \cdot \log(dt))$.
\end{enumerate}
\end{lemma}

Before proving Lemma~\ref{lem:query_complexity} we establish the following Lemma:
\begin{lemma}\label{lem:helper_query_complexity}
Assume the conditions of Lemma~\ref{lem:query_complexity}. Let $t' = 100t$, $\delta=1/(40000t(d+1)(d_\star+1))$, and $\alpha = 8 \cdot \ln(1/\delta)$.
The following three events happen with probability less than $\frac{1}{100}$ for each:

\begin{enumerate}
\item The total number of calls to Algorithm~\ref{O_V.alg.2} and Algorithm~\ref{alg:O_M.2} is bounded by $t'$.

\item The first $2t'$ vertices for which the operation $\lowest$ is called are $\alpha$-usual.

\item For the first $2t'$ vertices $v$ for which the operation $\lowest$ is called, the size of the set $T$ generated in the $j^{\rm th}$ execution of Step~\ref{step:generate_subset} of the operation is bounded by $\alpha \cdot \max\{1,\deg(v) \cdot 2^{j-d_\star}\}$.
\end{enumerate}

\end{lemma}
\BPF
For every group of parallel edges, the operation $\lowest$ lists only the edge with the lowest number. For the purpose of this analysis we assume that the operation lists in fact all occurences of a given parallel edge. The final complexity is only reduced because of the fact that some unnecessary calls are omitted.

\begin{enumerate}
\item Let us bound the probability that one of the above events does not occur.
By Markov's inequality the probability that the first event does not occur is bounded by $\frac{1}{100}$.
\item We shall now prove that the first $2t'$ vertices for which the operation $\lowest$ is called are $\alpha$-usual. The total number of vertices that have an incident edge for which the process generating random numbers is simulated in the above calls is bounded by $2t' \cdot (d+1)$.
The property of being $\alpha$-usual is a function of only random numbers assigned to incident edges. \\
For $\mathcal J_i$ let $X = \sum_{j=1}^{s} X_j$ where $p=\Pr[X_j = 1] = |\mathcal J_i|$, $s=\deg(v)$, i.e. $X$ is the number of all incident edges to $v$ with random numbers in $\mathcal J_i$.
From Lemma~\ref{lem:expectation_divergence} we get that:
\BEQN
\Pr \big[\sum_{i}X_i > \alpha \cdot \max\{1,|\mathcal J_i|\deg(v)\} \big] &=& \Pr \big[\sum_{i}X_i > 8 \cdot \ln(1/\delta) \cdot \max\{1,ps\}\big] \nonumber \\
&\leq&  \Pr[\sum_{i}X_i > 6 \cdot \ln(1/\delta) \cdot \max\{1,ps\}] < \delta \nonumber
\EEQN
Also, from Lemma~\ref{lem:expectation_divergence} we get that:
\BEQN
\Pr \big[\sum_{i}X_i < \frac{\deg(v)\cdot |\mathcal J_i|}{2}\big] &=& \Pr \big[\sum_{i}X_i < \frac{ps}{2} \big] < \delta \nonumber
\EEQN
i.e. $v$ is not $\alpha$-\emph{usual} because of $\mathcal J_i$ with probability less than $2\delta$. From union bound on all
all $i\in [d_\star +1]$ we get that vertex $v$ is not $\alpha$-\emph{usual} with probability less than $2\delta(d_\star +1)$.\\
Using the union bound again, this time over the vertices incident to edges for which the random process is run, the probability that any of them is not $\alpha$-usual is bounded by
\BEQ 2t' \cdot (d+1)\cdot 2\delta(d_\star +1) = 400t\delta(d+1)(d_\star+1) = \frac{1}{100}.\nonumber \EEQ

\item We need to prove that for the first $2t'$ vertices $v$ for which the operation $\lowest$ is called, the size of the set $T$ generated in the $j^{\rm th}$ execution of Step~\ref{step:generate_subset} of the operation is bounded by $\alpha \cdot \max\{1,\deg(v) \cdot 2^{j-d_\star}\}$. \\ Let $v$ be one of the first $2t'$ vertices for which the operation $\ds[v].\lowest$ is called. Observe that in the $j^{\rm th}$ iteration of the loop {\bf while},
$(\mbox{\tt next\_lb} - \mbox{\tt lb})/(1 - \mbox{\tt lb})$ is at most $2^{j-d_\star}$.
Therefore, it follows from Lemma~\ref{lem:expectation_divergence} that for each $j\in\{1,\ldots,d_\star+1\}$, the size of
the set $T$ in Algorithm~\ref{ds:lowest} selected in the $j^{\rm th}$ execution of Step~\ref{step:generate_subset} is bounded by
$\alpha \cdot \max\{1,\deg(v) \cdot 2^{j-d_\star}\}$ with probability $1-\delta$.
By the union bound over all $j$ and the first $2t'$ vertices, the probability that the third event does not occur is bounded by \BEQ 2t'(d_\star + 1)\delta = 200t(d_\star + 1)\cdot 1/(40000t(d+1)(d_\star+1) < \frac{1}{100} \nonumber \EEQ
\end{enumerate}
\EPF

Summarizing, the probability that at least one of the three events does not occur is bounded by
$$\frac{3}{100} < \frac{1}{20}$$

Let us now prove Lemma~\ref{lem:query_complexity} assuming that the events in Lemma~\ref{lem:helper_query_complexity} occur.

\medskip
\BPFOF{Lemma~\ref{lem:query_complexity}}
\begin{enumerate}
\item We need to prove that the total number of calls to Algorithms~\ref{O_V.alg.2} and~\ref{alg:O_M.2} is $O(t)$.
This follows directly from Lemma~\ref{lem:helper_query_complexity}, we proved it there for $t' = O(t)$.
\item We need to show that the operation $\lowest$ in data structures $\ds[v]$ runs at most $O(t)$ times. \\
The total number of vertices $v$ for which the operation $\ds[v].\lowest$ is called is bounded by $2t'$, because a call to one of the oracles (Algorithms~\ref{O_V.alg.2} and~\ref{alg:O_M.2}) requires calling the operation $\lowest$ for at most two vertices.
It follows from the implementation of the oracles that the operation $\ds[v].\lowest$ is executed at most $3t' = O(t)$ times if the number of oracle calls is bounded by $t'$ (which was proved in Lemma~\ref{lem:helper_query_complexity}). This is true because in Algorithm~\ref{O_V.alg.2} we call $\ds[v].\lowest$ once and in Algorithm~\ref{alg:O_M.2} we call $\ds[v].\lowest$ twice.

\item We will now show that the query complexity of $\mathcal A$ is $O(t \cdot \log^2 (dt))$.\\
For each vertex $v$, denote $k_v\in [0, \deg(v)]$ the number of times we call $\ds[v].\lowest(k)$ on $v$.
We assume that if the operation is not executed for a given vertex, then $k_v = 0$.
It holds that:
$$\sum_{v \in V} k_v \le 3t'$$
We now attempt to bound the query complexity necessary to execute the operation $\ds[v].\lowest$ for a given $v$ such that $k_v > 0$. Note that the expected number of edges with random numbers in a given $\mathcal J_i$ is $\deg(v)/2^{i-1}$.
Recall that from Lemma~\ref{lem:helper_query_complexity} we know that the first $2t'$ vertices for which the operation $\lowest$ is called are $\alpha$-usual. From the lower bound of $\alpha$-usual (Definition~\ref{def:alpha_usual}) we get that
If $\deg(v) \cdot |\mathcal J_i| \ge \alpha$, then the number of edges with random numbers in $\mathcal J_i$ is at least
\BEQ \deg(v) \cdot |\mathcal J_i| / 2. \nonumber \EEQ
Therefore, if
$$\deg(v) \cdot |\mathcal J_i| = \deg(v)/2^{i-1} \ge \max\{2\alpha,2k_v\}$$
then the number of edges with random numbers in $\mathcal J_i$ is at least
$$\frac{\max\{2\alpha,2k_v\}}{2} = \max\{\alpha,k_v\} \ge k_v $$
i.e. if $i$ is such that $\deg(v)/2^{i-1}$ is at least $\max\{2\alpha,2k_v\}$, then at least
$k_v$ edges incident to $v$ have random numbers in $\mathcal J_i$. This also holds
for $i$ such that $\deg(v)/2^{i-1} \ge 2\alpha k_v$. Let $i_v$ be the largest integer $i$
such that $2^i\le \frac{\deg(v)}{\alpha k_v}$ (remember $i = d_{\star + 1}, d_{\star} \cdots$).
Since $i_v$ is the maximum $i$ that satisfies this, then
$$2^{i_v+1} > \frac{\deg(v)}{\alpha k_v} \Rightarrow 2^{i_v} > \frac{\deg(v)}{2\alpha k_v} \Rightarrow 2^{-i_v} < \frac{2\alpha k_v}{\deg(v)}$$
The body of the loop {\bf while} in Algorithm~\ref{ds:lowest} is executed at most $d_\star + 2 - i_v$ times for $v$ (remember we start from $i=d_{\star+1}$), independently of how many times the operation is executed for $v$, because all relevant edges incident to $v$ are discovered during these iterations. From Lemma~\ref{lem:helper_query_complexity} we know that the size of the set $T$ in Algorithm~\ref{ds:lowest} selected in the $j^{\rm th}$ execution of this loop is bounded by
$\alpha \cdot \max\{1,\deg(v) \cdot 2^{j-d_\star}\}$.
Furthermore, the sum of sizes of all sets $T$ generated for $v$ is bounded by
\begin{eqnarray*}
\sum_{j=1}^{d_\star+2-i_v} \alpha\cdot\max\{1,\deg(v) \cdot 2^{j-d_\star}\}
&\le&
\alpha (d_\star +1) + 2\alpha \cdot \deg(v) \cdot 2^{2-i_v}\\
&\le& \alpha (d_\star +1) + 16\alpha^2 k_v.
\end{eqnarray*}
This also bounds the number of neighbor queries for $v$.
Since these are the only neighbor queries in the algorithm,
by summing over all $v$ with $k_v \ge 0$, the total number of neighbor queries is bounded by
$$2t' \cdot \alpha (d_\star +1) + \sum_{v \in V} 16\alpha^2 k_v \le
200\alpha t (d_\star + 1) + 16\alpha^2 \cdot 300 t = O(\alpha t (d_\star + \alpha)) = O(t \cdot \log^2(dt)).$$
(Recall $t'=200t$ and that $\sum_{v \in V} k_v \le 3t'$).
Note that degree queries appear only in Step~\ref{step:generate_subset} of the operation $\ds[v].\lowest$
with one query to discover the size of the set from which a subset is selected. The number of
degree queries is in this case bounded by the total number of executions of Step~\ref{step:generate_subset}, which is at most $O(t \cdot \log d)$. Summarizing, the total query complexity is $O(t \cdot \log^2(dt))$.

\item Finally, we need to prove that the total toll for running Step~\ref{step:generate_subset} of Algorithm~\ref{ds:lowest} is $O(t \cdot \log(dt))$.
Recall that the toll is defined as $kp$ where $k$ is the number of neighbors and $p$ is the probability to selecting each of them in an execution of Step~\ref{step:generate_subset} of Algorithm~\ref{ds:lowest}.
Using arguments as above, the toll for running Step~\ref{step:generate_subset} in the operation $\ds[v].\lowest$ for a given $v$ is bounded by
$$
\sum_{j=1}^{d_\star+2-i_v} \deg(v) \cdot 2^{j-d_\star} \le 2 \cdot \deg(v) \cdot 2^{2-i_v} \le 8 \cdot \deg(v) \cdot \frac{2 \alpha k_v}{\deg(v)} = 16\alpha k_v$$
By summing over all vertices $v$, we obtain a bound on the total toll:
$$\sum_{v\in V} 16\alpha k_v \le 4800 \alpha t = O(t \cdot \log (dt)).$$
\end{enumerate}
\EPFOF

\subsection{Efficient Implementation}
\label{sec:efficient_implementation}

We have already introduced techniques that can be used to show an approximation algorithm
whose query complexity has near-linear dependence on the maximum degree $d$.
Unfortunately, a straightforward implementation
of the algorithm results in a running time with approximately quadratic dependence on $d$. The goal of this section is to remove a factor of approximately $d$ from the running time of the algorithm.
Our main problem is how to efficiently simulate Step~\ref{step:generate_subset} in the operation $\lowest$. Note that Step~\ref{step:generate_subset} is sampling from a binomial distribution.\\
First, in Lemma~\ref{lem:binomial_generation}, we prove that there is an algorithm that can simulate a binomial distribution which runs in efficient time.
Finally, in Theorem~\ref{thm:running_time}, we will show how to use it in our algorithms and how to bound the running time by $O(t\cdot\log^3(dt))$.

\noindent
We start by defining the binomial distribution.

\begin{definition}
We write $B(k,p)$, where $k$ is a positive integer and $p \in [0,1]$, to denote the \emph{binomial distribution with success probability $p$} on $\{0,1,\ldots,k\}$ distributed as $\sum_{i=1}^k X_i$, where each $X_i$, $1 \le i \le k$, is an independent random variable that equals $1$ with probability $p$, and $0$ with probability $1-p$.
\end{definition}

It is well known that the probability that a value drawn from the binomial distribution $B(k,p)$ equals $q$ is $\binom{k}{q}p^q(1-p)^{k-q}$. We now show how to efficiently sample from this distribution.

\begin{lemma}\label{lem:binomial_generation}
Let $a$, $b$, $k$, and $Q$ be positive integers, where $a \le b$ and $Q > 1$,
that can be represented in the standard binary form, using a constant number of machine words.
There is an algorithm that takes $a$, $b$, $k$, and $Q$ as parameters, runs in $O(\max\{
ka/b,1\} \cdot \log Q)$ time, and outputs an integer selected from a distribution $\mathcal D$ on $\{0,1,\ldots,k\}$ such that the total variation distance between $\mathcal D$ and $B(k,a/b)$ is bounded by $1/Q$.
\end{lemma}

\BPF
If $a=b$, then the algorithm can return the trivial answer in $O(1)$ time, so we can safely assume for
the rest of the proof that $a < b$. Let $p=a/b$ and let $q_i = \binom{k}{i}p^i(1-p)^{k-i}$ be the probability of drawing $i$ from $B(k,p)$. Let $s = \min\{ 6 \cdot \ln(2Q) \cdot \max\{1,ka/b\}, k\}$. For each $i \le s$, we compute a real number $q'_i \in [0,1]$ such that $q_i - q'_i \le 1/2(k+1)Q$ and $\sum_{i=0}^s q'_i = 1$ (details about how to compute those $q_i'$ are given in Lemma~\ref{lem:compute_q_i'}).
Then we select the output of the algorithm from the distribution given by $q'_i$'s.
We write $\mathcal D$ to denote this distribution.

Let us bound the total variation distance between this distribution and $B(k,p)$.
It suffices to show that for every subsets $S$ of $\{0,\ldots,k\}$, the probability
of selecting an integer from $S$ in $B(k,p)$ is not greater by more than $1/Q$,
compared to the probability of selecting an integer in $S$ from $\mathcal D$.
Consider an arbitrary such set $S$.
Let $S_1$ be the subset of $S$ consisting of numbers at most $s$.
Let $S_2$ be the subset of $S$ consisting of integers greater than $s$.
We have
\BEQ \sum_{i \in S} q'_i \ge \sum_{i \in S_1} q'_i\ge \sum_{i \in S_1} \left (q_i - \frac{1}{2(k+1)Q}\right) \ge \left(\sum_{i \in S_1} q_i\right) - \frac{1}{2Q}. \label{fact:S_1}\EEQ
Recall that $X_i = 1$ with probability $p$. If $s=k$ then
$$\Pr \big[\sum_{i=1}^kX_i > s \big] = \big[\sum_{i=1}^kX_i > k \big] = 0$$
If $s=6 \cdot \ln(2Q) \cdot \max\{1,ka/b\}$ then we define $\delta = \frac{1}{2Q}$, and from Lemma~\ref{lem:expectation_divergence} we have that
$$\Pr \big[\sum_{i=1}^k > 6 \cdot \ln(\frac{1}{\delta}) \cdot \max\{1,pk\} \big] < \delta$$
Hence,
$$\Pr \big[\sum_{i=1}^k > s \big] < \frac{1}{2Q}$$
In other words: the probability that a number greater than $s$ is being selected from $B(k,p)$ (i.e. s $X_i$'s are 1) is bounded by $\frac{1}{2Q}$.
Therefore,
\BEQ\left(\sum_{i \in S_2} q_i\right) < \frac{1}{2Q} \label{fact:S_2} \EEQ
From~\ref{fact:S_1} and~\ref{fact:S_2} we get:
$$\sum_{i \in S} q'_i \ge \left(\sum_{i \in S_1} q_i\right) - \frac{1}{2Q} +
\left(\sum_{i \in S_2} q_i\right) - \frac{1}{2Q} \ge
\left(\sum_{i \in S} q_i\right) - \frac{1}{Q},$$
Therefore,
$$\sum_{i \in S} [q_i - q_i'] \leq \frac{1}{Q},$$
which proves our Lemma.
Next, in Lemma~\ref{lem:compute_q_i'} we will also show that the running time of the algorithm is $O(s) = O(\max\{k\frac{a}{b}, 1\}\log(Q))$.
\EPF

We now describe how to compute values $q_i'$ that are approximation to $q_i$.

\begin{lemma}\label{lem:compute_q_i'}
Recall: $a < b$, $p=a/b$ and $q_i = \binom{k}{i}p^i(1-p)^{k-i}$ (probability of drawing $i$ from $B(k,p)$). Let $s = \min\{ 6 \cdot \ln(2Q) \cdot \max\{1,ka/b\}, k\}$. For each $i \le s$, we can compute a real number $q'_i \in [0,1]$ such that $q_i - q'_i \le 1/2(k+1)Q$ and $\sum_{i=0}^s q'_i = 1$. The total running time is $O(\max\{ka/b,1\} \cdot \log Q)$.
\end{lemma}

\BPF
Observe that for $1 \le i \le k$:
$$q_i = q_{i-1} \cdot \frac{k+1-i}{i} \cdot \frac{p}{1-p}$$
Let $t_i = \frac{q_i}{q_0}$ for $0 \le i \le s$. It holds that for $1 \le i \le s$:
\begin{equation}\label{eqn:binomial_generation}
t_i = t_{i-1} \cdot \frac{k+1-i}{i} \cdot \frac{p}{1-p} = t_{i-1} \cdot \frac{k+1-i}{i} \cdot \frac{a}{b-a}
\end{equation}
Note that for $0 \le i \le s$:
\BEQ \frac{t_i}{\sum_{j \le s} t_j} = \frac{\frac{q_i}{q_0}}{\frac{1}{q_0}{\sum_{j \le s} q_j}} \ge q_i \label{fact:t_i_ge}\EEQ
Suppose now that instead of $t_i$, we use $t'_i \ge 0$, $0 \le i \le s$, such that $|t_i - t'_i| \le \frac{\max_{0 \le j \le s} t_j}{4(k+1)^2Q}$.
Then from the definition of $t_i'$ we get:
\BEQ t_i' \ge t_i - \frac{\max_{0 \le j \le s} t_j}{4(k+1)^2Q} \EEQ
Also:
\BEQ \sum_{j \le s} t'_j \le s \cdot \frac{(\max_{0 \le j \le s} t_j)}{4(k+1)^2Q + \sum_{j \le s} t_j} \le
\frac{\max_{0 \le j \le s} t_j}{4(k+1)Q + \sum_{j \le s} t_j} \le \big(1+ \frac{1}{4(k+1)Q}\big) \cdot \sum_{j \le s} t_j\EEQ
We have
\begin{eqnarray*}
\frac{t'_i}{\sum_{j \le s} t'_j}
&\ge& \frac{t_i - \frac{\max_{0 \le j \le s} t_j}{4(k+1)^2Q}}{(1+\frac{1}{4(k+1)Q})\sum_{j \le s} t_j}\\
\mbox{ (From~\ref{fact:t_i_ge} and that $\max_{0 \le j \le s} t_j \leq \sum_{j \le s} t_j$)} &\ge& \frac{q_i\cdot \sum_{j \le s} t_j}{(1+\frac{1}{4(k+1)Q})\cdot \sum_{j \le s} t_j} - \frac{\frac{\sum_{j \le s} t_j}{4(k+1)^2Q}}{{(1+\frac{1}{4(k+1)Q})\cdot \sum_{j \le s} t_j}} \\
\mbox{ ( Since $1 + \frac{1}{4(k+1)^2Q} \ge 1$)} &\ge& \frac{q_i}{(1+\frac{1}{4(k+1)Q})} - \frac{1}{4(k+1)^2Q} \\
&\ge& q_i \big(1-\frac{1}{4(k+1)Q}\big) - \frac{1}{4(k+1)^2Q}\\
&\ge& q_i - \frac{1}{4(k+1)Q} - \frac{1}{4(k+1)^2Q}
\ge q_i - \frac{1}{2(k+1)Q}.
\end{eqnarray*}
So eventually we get that
\BEQ q_i - \frac{t'_i}{\sum_{j \le s} t'_j} \le \frac{1}{2(k+1)Q} \EEQ
Also, note that $\sum_{i=0}^s\frac{t'_i}{\sum_{j=0}^s t'_j} = 1$.\\
Therefore, in our distribution $\mathcal D$, we will define $q_i' = \frac{t'_i}{\sum_{j \le s} t'_j}$.\\
 It remains to show how we obtain $t'_i$ with the desired properties.
For this purpose, we use floating-point arithmetic.
Each positive number that we obtain during the computation
is stored as a pair $\langle S,E\rangle$ representing $S \cdot 2^E$.
We require that $2^\alpha \le S < 2^{\alpha+1}$ and $|E| \le \beta$, for some $\alpha$ and $\beta$ to be set later. If we can perform all standard operations on these integers in $O(1)$ time, then we can perform the operations on the represented positive real numbers in $O(1)$ time as well.
We call $S$ a \emph{significand} and $E$ an \emph{exponent}.

In particular, to multiply two numbers $\langle S_1,E_1 \rangle$ and $\langle S_2,E_2 \rangle$ it suffices to multiply $S_1$ and $S_2$, truncate the least significant bits of the product,
and set the new exponent accordingly.
If these two numbers are multiplicative $(1\pm \delta_1)$- and $(1\pm\delta_2)$-approximations to some quantities $X_1$ and $X_2$, respectively, then
the product of $S_1$ and $S_2$ in our arithmetic is a multiplicative $(1\pm(\delta_1+\delta_2+\delta_1\delta_2+2^{-\alpha}))$-approximation to $X_1 X_2$. If $\delta_1 < 1$, then the product is a $(1\pm(\delta_1+2\delta_2+2^{-\alpha}))$-approximation.

For each $i$ of interest, one can easily compute a multiplicative $(1\pm C \cdot 2^{-\alpha})$-approximation for $\frac{k+1-i}{i} \cdot \frac{a}{b-a}$ in our arithmetic, where $C > 1$ is a constant.
We make the assumption that $3Ck2^{\alpha} \le 1$, which we satisfy later by setting a sufficiently large $\alpha$. Hence we use Equation~\ref{eqn:binomial_generation} to obtain a sequence of multiplicative $(1\pm3Ck2^{-\alpha})$-approximations $t'_i$ for $t_i$, where $0 \le i \le s$.
At the end, we find the maximum $t'_i$, which is represented as a pair $\langle S_i,E_i\rangle$.
For all other $t'_i$, we no longer require that $S_i \ge 2^\alpha$ and
we modify their representation $\langle S_i,E_i\rangle$ so that $E_i$ is the same as in the representation of the maximum $t'_i$.
In the process we may lose least significant bits of the some $t'_i$ or even all non-zero bits. Assuming again that $3Ck2^{-\alpha} < 1$,
the maximum additive error $|t_i - t'_i|$ we get for each $i$ for the modified representation is bounded by
$$3Ck2^{-\alpha}\cdot t_i + 2^{-\alpha} \cdot \max_j t'_j \le
3Ck2^{-\alpha}\cdot t_i + 2 \cdot 2^{-\alpha} \cdot \max_j t_j
\le (3Ck+2)\cdot 2^{-\alpha} \cdot \max_j t_j,$$
where the first error term comes from the multiplicative error we obtain approximating each $t_i$
and the second error term comes from making all exponents in the representation match the exponent of the largest $t'_i$. Finally, we set $\alpha = \lceil \log((3Ck+2) \cdot 4 (k+1)^2Q)\rceil$.
This meets the previous assumption that $3Ck2^{-\alpha} < 1$ and the guarantee on the error we may make on each $t'_i$ is as desired.
Note that since $k$ and $Q$ can be represented using a constant number of words, so can integers of size at most $2^{\alpha+1}$. To bound $\beta$, observe that every $\frac{k+1-i}{i} \cdot \frac{a}{b-a}$
lies in the range $[1/kb,kb]$, which implies that
all $t_i$ lie in $[1/kb^k,kb^k]$, and the maximum absolute value of an exponent we need is of order $O(k \log (kb))$, which can be stored using a constant number of machine words.

To generate a random number from $\mathcal D$, we consider only the significands $S_i$ in the final modified representation of $t'_i$'s, and select each $i$ with probability $S_i/\sum_{j < s} S_j
= t'_i/\sum_{j < s} t'_j$. The total running time of the algorithm is $O(s)$.
\EPF

We are ready to prove that the entire algorithm can be implemented efficiently. 
We use the algorithm of Lemma~\ref{lem:binomial_generation} for efficiently 
simulating Step~\ref{step:generate_subset} in the operation $\lowest$.
\ifnum\conf=1
We next restate Theorem~\ref{thm:running_time} and prove it.

\bigskip\noindent{\bf Theorem~\ref{thm:running_time} (Restated)}~
{\it
Consider an algorithm $\mathcal A$ that queries the input graph only via the oracle described as Algorithm~\ref{O_V.fig}. Let $t \ge 1$ be a bound on the expected resulting number of calls in $\mathcal A$ to the oracles described as Algorithm~\ref{O_V.fig} and Algorithm~\ref{alg:O_M}, and such that $t$ fits into a constant number of machine words using the standard binary representation. Let $d$ be an upper bound on the maximum degree of the input graph.

Suppose that calls to Algorithm~\ref{O_V.fig} are replaced with calls to Algorithm~\ref{O_V.alg.2}.
The oracles described as Algorithm~\ref{O_V.alg.2} and Algorithm~\ref{alg:O_M.2} can be implemented in such a way that with probability $4/5$ all of the following events hold:
\begin{itemize}
\item The number of queries to the graph is $O(t \cdot \log^2 (dt))$.
\item The total time necessary to compute the answers for the queries to the oracles is $O(t \cdot \log^3 (dt))$.
\item The distribution of the answers that the oracle gives is $\mathcal D$ such that
for some other distribution $\mathcal D'$ over answers, the convex combination $\frac{4}{5}\cdot\mathcal D + \frac{1}{5}\cdot \mathcal D'$ is the distribution of answers of the oracle described as Algorithm~\ref{O_V.fig}.
\end{itemize}
}

\medskip
\else
\BT \label{thm:running_time}
Consider an algorithm $\mathcal A$ that queries the input graph only via the oracle described as Algorithm~\ref{O_V.fig}. Let $t \ge 1$ be a bound on the expected resulting number of calls in $\mathcal A$ to the oracles described as Algorithm~\ref{O_V.fig} and Algorithm~\ref{alg:O_M}, and such that $t$ fits into a constant number of machine words using the standard binary representation. Let $d$ be an upper bound on the maximum degree of the input graph.

Suppose that calls to Algorithm~\ref{O_V.fig} are replaced with calls to Algorithm~\ref{O_V.alg.2}.
The oracles described as Algorithm~\ref{O_V.alg.2} and Algorithm~\ref{alg:O_M.2} can be implemented in such a way that with probability $4/5$ all of the following events hold:
\begin{itemize}
\item The number of queries to the graph is $O(t \cdot \log^2 (dt))$.
\item The total time necessary to compute the answers for the queries to the oracles is $O(t \cdot \log^3 (dt))$.
\item The distribution of the answers that the oracle gives is $\mathcal D$ such that
for some other distribution $\mathcal D'$ over answers, the convex combination $\frac{4}{5}\cdot\mathcal D + \frac{1}{5}\cdot \mathcal D'$ is the distribution of answers of the oracle described as Algorithm~\ref{O_V.fig}.
\end{itemize}
\ET
\fi

\BPF
Let $a_\star = d \cdot O(t)$, where $O(t)$ is the
bound from Lemma~\ref{lem:query_complexity}
on the number of vertices for which the operation $\lowest$ is called.
If the event specified in Lemma~\ref{lem:query_complexity} occurs, then
$a_\star$ is an upper bound on the number of edges for which the process for generating random
numbers is simulated. Let $b_\star = O(t) \cdot (d_\star + 1) = O(t \log d)$, where $O(t)$ is the same bound as above. Then $b_\star$ bounds the number of times Step~\ref{step:generate_subset} in Algorithm~\ref{ds:lowest} is run, provided the event specified in Lemma~\ref{lem:query_complexity} occurs. Let $Q = 20 b_\star$.

Let $c_\star = \max\{d_\star, \lceil\log(20a^2_\star)\rceil\}$. Since it is impossible to
generate and store real numbers, we assign to edges uniform random numbers from the set $\{i/2^{c_\star}: 1 \le i \le 2^{c_\star}\}$, instead of the set $(0,1]$.
This can be seen
as selecting a random number from $(0,1]$ and then rounding it up to the next multiplicity of
$1/2^{c_\star}$. In particular, for every $i \in \{1,\ldots,2^{c_\star}\}$, all numbers in $((i-1)/2^{c_\star},i/2^{c_\star}]$ become $i/2^{c_\star}$. Observe also that each range $\mathcal I_j$ is a union of some number of sets $((i-1)/2^{c_\star},i/2^{c_\star}]$, because $c_\star \ge d_\star$. This means that there is no need to modify the process for generating random numbers, except for selecting a random $i/2^{c_\star}$ in a specific $\mathcal I_j$, instead of an arbitrary real number from $\mathcal I_j$. Observe also that as long we do not select the same number $i/2^{c_\star}$ twice, the entire exploration procedure behaves in the same way as in the idealized algorithm, since the ordering of numbers remains the same.

Note that due to the assumption in the lemma statement, $t$ can be represented in the standard binary form, using a constant number of machine words. This is also the case for $d$, because of the standard assumption that we can address all neighbors of all vertices in neighbor queries.
This implies that $Q = O(t \log d)$ also has this property. Finally, the probabilities $\frac{\mbox{\tt next\_lb} - \mbox{\tt lb}}{1 - \mbox{\tt lb}}$ can easily be expressed
using fractions $a/b$, where $a$ and $b$ are of order $O(d)$, and therefore, fit into a constant number of machine words as well. This implies that we can use the algorithm of Lemma~\ref{lem:binomial_generation}. Instead of directly simulating Step~\ref{step:generate_subset}, we proceed as follows. First, we run the algorithm of Lemma~\ref{lem:binomial_generation} with the error parameter $Q$ to select a number $t$ of edges in $T$. Then we select a random subset of edges of size $t$. This can be done in $O(t \log d)$ time.

We show that the algorithms and data structures can be implemented in such a way that the main claim of the theorem holds, provided the following events occur:
\begin{itemize}
\item the events described in the statement of Lemma~\ref{lem:query_complexity},
\item the rounded numbers assigned to the first $a_\star$ edges for which the process for generating random numbers is simulated are different,
\item the first $b_\star$ simulations of the algorithm described by Lemma~\ref{lem:binomial_generation} do not result in selecting a random number from the part on which the output distribution of the algorithm and the binomial distribution differ.
\end{itemize}

The first of the events does not happen with probability at most 1/10. This follows from Lemma~\ref{lem:query_complexity}. Consider the second event. The probability that two random numbers
$i/2^{c_\star}$ are identical is bounded by $1/2^{c_\star}\le 1/(20 a_\star^2)$. Consider the
first $a_\star$ edges for which the process generating random numbers is run. The expected number of pairs of the edges that have the same random number is bounded by $a_\star^2 \cdot 1/(20 a_\star^2)= 1/20$. By Markov's inequality, the probability that two of the edges
have the same random number assigned is bounded by $1/20$. Finally, the probability that the last event does not occur is bounded by $1/20$ as well via the union bound. Summarizing, the events occur with probability at least $4/5$.

We now bound the running time, provided the above events occur.
We assume that we use a standard data structure (say, balanced binary search trees) to maintain
collections of items. The time necessary for each operation in these data structures is of order at most the logarithm of the maximum collection size. For instance, we keep a collection of data structures $\ds[v]$ for $v$ that appear in our algorithm. We create $\ds[v]$ for a given $v$ only when it is accessed for the first time. Observe that the number of $v$ for which we have to create $\ds[v]$ is bounded by the query complexity $O(t \log^2(dt))$, because of how we access vertices.
Therefore, accessing each $\ds[v]$ requires at most $O(\tau)$ time, where
we write $\tau$ to denote the logarithm of the bound on the query complexity.
That is, $\tau = O(\log t + \log\log d)$.

The time necessary to run Algorithm~\ref{O_V.alg.2} is bounded by $O(\tau)$, which we need to locate the data structure $\ds[v]$
for a given $v$, plus $O(1)$ time per each call to Algorithm~\ref{alg:O_M.2} (we do not include the cost of running Algorithm~\ref{alg:O_M.2} or the operation $\lowest$ here; they are analyzed later). The amount of computation in Algorithm~\ref{O_V.alg.2} without the resulting calls to other procedures is bounded by $O(t \cdot \tau)$.

Consider now Algorithm~\ref{alg:O_M.2}. In every run, we first spend $O(\log t)$ time to check if we have already computed the answer for a given edge. Then locating the data structures $\ds[u]$ and $\ds[v]$ for the endpoints $u$ and $v$ costs at most $O(\tau)$. The running time of the reminder of the algorithm requires time proportional to the number of recursive calls. Therefore, the total amount of time spent executing Algorithm~\ref{alg:O_M.2} (without calls to other procedures) is bounded by $O(t \cdot \tau)$.

We now bound the running time necessary to execute all operations of data structures $\ds$.
The initialization of $\ds[v]$ (Algorithm~\ref{ds:init}) for a given $v$ can be done $O(1)$ time plus $O(\tau)$ time necessary
for inserting the data structure into the collection of all $\ds[v]$.
Overall, since at most $O(t \log^2(dt))$ data structures are created, the total time necessary to initialize the data structures $\ds[v]$ is $O(t \cdot \log^2(dt) \cdot \tau)$.
Setting a value for some edge in Algorithm~\ref{ds:setvalue} takes at most $O(\log d)$ time to insert the value into the mapping. This operation is run at most once for every neighbor query, so the total amount of computation in this procedure is $O(t \cdot \log^2(dt) \cdot \log d)$.
So far, the total computation time is bounded by $O(t \log^3(dt))$.

Clearly, running the operation described by Algorithm~\ref{ds:lower_bound} takes $O(1)$ time,
so overall the total amount of computation in all executions of Algorithm~\ref{ds:lower_bound} is not greater than some constant times the total amount of computation in the operation $\lowest$ (Algorithm~\ref{ds:lowest}). Hence it suffices to bound the total amount of computation in
Algorithm~\ref{ds:lowest}, which we do next.

Recall that Algorithm~\ref{ds:lowest} is run at most $O(t)$ times.
Therefore all operations in the loop {\bf while} are run at most $O(t \log d)$ times. The total size of sets $S$ in Step~2 is bounded by the query complexity, and discovering each element of $S$ costs at most $O(\log d)$ time, if the data structure $\mbox{\rm assigned\_number}$ is properly implemented, using augmented balanced binary search trees. Therefore the total cost of running Step~2 is at most $O(t \cdot  \log d + t \cdot \log^2(dt) \cdot \log d) = O(t \cdot \log^2(dt) \cdot \log d)$. In Step~3, we use the algorithm of Lemma~\ref{lem:binomial_generation}. The total toll for running the algorithm is $O(t \cdot \log(dt))$. Therefore, the total time necessary to simulate all executions of Step~2 is bounded by $O((t \cdot \log d + t \cdot \log(dt)) \cdot \log Q) =  O(t \cdot \log^2(dt))$. The total number of executions of the body of the loop {\bf foreach} in Step~4 is bounded by the query complexity $O(t \cdot \log^2(dt))$ times 2. The time required to execute the body of the loop is dominated by the following two kinds of operations.
One kind is querying and modifying the data structure $\mbox{\tt assigned\_number}[w]$ and the data structure for $S$. With a proper implementation (say, augmented balanced binary search trees) these operations take at most $O(\log d)$ time each. The other kind of operation is
locating $\ds[w]$ for the discovered neighbor $w$, which takes most $O(\tau)$ time.
The total computation time for all executions of the loop {\bf foreach} is therefore bounded by
$O(t \cdot \log^3(dt))$.

Finally sorting $S$ never takes more than $O(|S| \log d)$ time, because $|S| \le d$,
and each element of $S$ can be added at the end of the list $\mbox{\tt sorted}$ in amortized $O(1)$
time if the list is implemented using extendable arrays. This amounts to $O(t \cdot \log^2(dt) \cdot \log d)$ in all executions of Step~11. At the end of the operation, the requested $k^{\rm th}$ adjacent edge can be returned in $O(1)$ time.

Summarizing, the computation of the answers of the oracles takes at most $O(t \cdot \log^3(dt))$ time, if all the desired events occur, which happens with probability at least $4/5$. Note that when these events occur, then also despite rounding random numbers assigned to edges, the implementation does not diverge from the behavior of the idealized oracle.
\EPF

%% file: near-opt.tex
\section{The Near-Optimal Algorithms}\label{near-opt.sec}

\ifnum\conf=1
By combining Theorem~\ref{avg-calls.thm} with 
Theorem~\ref{thm:running_time}
 we can obtain an algorithm whose query complexity
and running time are $O\left(\frac{\rho\cdot \bd}{\eps^2}\cdot \log^3(d/\eps)\right)$.
In general this bound may be $O\left(\frac{d^2}{\eps^2}\cdot \log^3(d/\eps)\right)$.
By running slightly modified versions of the algorithm on certain modified versions
of the graph we can get the following improved results for obtaining a $(2,\eps)$-estimate
of the size of a minimum vertex cover (with constant success probability):
\BI
\item An algorithm with query complexity 
 $O(\frac{d}{\eps^3} \cdot \log^3\frac{d}{\eps})$ 
and running time $O(\frac{d}{\eps^3} \cdot \log^3\frac{d}{\eps})$,

\item An algorithm with query complexity 
$O(\frac{\bd}{\eps^4} \cdot \log^2\frac{\bd}{\eps})$
and running time 
$O(\frac{\bd}{\eps^4} \cdot \log^3\frac{\bd}{\eps})$.
\EI
The idea behind the first result is to add, for each vertex $v$ in the graph $G$, a
``shadow'' vertex $v'$. Between $v$ and $v'$ there are $\Theta(\eps d)$ parallel
edges, and there are $\Theta(d)$ self-loops incident to $v'$. Let the resulting
graph be denoted $\tG$. The maximum degree
in $\tG$ is $O(d)$, the average degree is $O(d)$, and the ratio,
$\rho$, between the maximum and the minimum degree is $1/\eps$. 
The main observation is that if we select a random ranking
of the edges in $\tG$, then with high probability, the following holds 
for all but an $(\eps/4)$-fraction of the vertices $v$: The edge with minimal
rank among the parallel edges between $v$ and $v'$ and the self-loops incident to $v'$,
is a self loop, in which case it is added to the matching, while none of the parallel
edges between $v$ and $v'$ are added to the matching.
If the ranking has this property, then the number of vertices 
in the original graph $G$ that belong to the vertex cover of $\tG$ 
(which is determined by the ranking)
is at most $2\VCopt(G) + (\eps/4)n$. If we obtain a good estimate on the number
of 
these vertices (by sampling $\Theta(1/\eps^2)$ vertices in $G$), 
then we can obtain a $(2,\eps)$-estimate of $\VCopt(G)$.
By applying Theorem~\ref{avg-calls.thm} (to $\tG$), we get an upper 
bound of $O(d/\eps)$ on the number of recursive calls, and we can also show that
our efficient implementation can be applied to to $\tG$.

To obtain the second result, we combine the above type of construction with
an idea presented in~\cite{PR}: Since the fraction of vertices with degree 
 $8\bd/\eps$ is at most $\eps/8$, we can ``afford'' to add them to
the cover. The benefit is that the graph induced on the remaining vertices, 
denoted $\bG$, has a maximum degree of $8\bd/\eps$. By adding shadow vertices to deal with
the low-degree vertices in $\bG$ as described above, and additional shadow vertices to
deal with edges between vertices in $\bG$ and the high degree vertices in $G$
that are not in $\bG$, we get the second result.

\paragraph{Dense graphs.}
As noted in the introduction, while the focus of this paper is on algorithms that
access the graph through neighbor queries, we also consider a query model, more
approriate for dense graphs (in which $\bd = \Theta(n)$), that
allows vertex-pair queries (that is, queries of then form: ``is there an edge
between $u$ and $v$''). We can adapt our algorithm that uses neighbor queries
to an algorithm that uses vertex-pair queries, where the algorithm outputs
(with high constant probability) a $(2,\eps)$-estimate of the minimum vertex cover size, 
and has complexity $\tilde{O}(n/\eps^4)$ (which is almost linear in $\bd$ since
the graph is dense).

\medskip\noindent
For further details see 
\ifnum\hideappendix=1
the full paper.
\else
Appendix~\ref{near-opt.app}.
\fi
\fi

\def\NearOpt{
Theorem~\ref{avg-calls.thm} gives a bound on the expected number of recursive calls to oracles,
sufficient to compute an answer when the vertex cover oracle is called for a random vertex.
The expected number of calls is $O(\rho\cdot \bd)$,
where $\rho$ is the ratio between
the maximum degree $d$ and the minimum degree $d_{\min}$, and $\bd$ is the 
average degree.
(Recall that we assume without loss of generality that $d_{\min} \geq 1$.
For isolated vertices, the oracle answers that they are not in the vertex cover
in $O(1)$ time, and therefore, it suffices to focus on the subgraph consisting of non-isolated vertices.)

A straightforward application of Theorem~\ref{avg-calls.thm} gives a bound of $O(d^2)$ 
for graphs with maximum degree bounded by $d$. We show a bound of $O(d/\eps)$ for 
a \emph{modified} graph, which is preferable if $1/\eps < d$, and we also show how to 
use the modified graph to obtain an estimate for the minimum vertex cover size in the 
original input graph. We combine the obtained bound with 
Theorem~\ref{thm:running_time} to get a fast and query-efficient algorithm.

Next we show how to obtain an efficient algorithm for the case when only the average degree
of the input graph is bounded. Finally, we show how to adapt the algorithm to the dense graph case, when only vertex-pair queries are allowed.


\subsection{Bounded Maximum Degree}\label{bound-deg.subsec}

As we have mentioned above, we can assume that $1/\eps < d$.
We transform our graph into one with large minimum degree,
so that the ratio of maximum to minimum degree is small.
For a given graph $G = (V,E)$ with maximum degree $d$, 
consider a graph $\tG = (\tV,\tE)$,
such that $\tV = V \cup V'$ and $\tE = E\cup E'$
where $V'$ and $E'$ are defined as follows.
The set $V'$ contains a ``shadow'' vertex $v'$ for each vertex $v\in V$,
and $E'$ contains $\lfloor \eps d \rfloor$ parallel edges between 
$v$ and $v'$, and $8d$ parallel self-loops for $v'$.

For a random ranking $\tilde{\pi}$ over $\tE$, 
for the output vertex cover $\C^{\tilde{\pi}}(\tG)$ on  the new graph $\tG$,
we are interested in bounding  the size of 
$\C^{\tilde{\pi}}(\tG) \cap V$
as compared to $\VCopt(G)$ (the size of a minimum vertex cover of $G$).
Since $\C^{\tilde{\pi}}(\tG) \cap V$ is a vertex cover of $G$,
we have that $|\C^{\tilde{\pi}}(\tG) \cap V| \geq \VCopt(G) $,
and so we focus on an upper bound for $|\C^{\tilde{\pi}}(\tG) \cap V|$.

Let $\tilde F$ be the set of all parallel edges connecting each $v$ with the corresponding $v'$.
By the properties of the construction of 
$\C^{\tilde{\pi}}(\tG) \cap V$,
we have 
$$|\C^{\tilde{\pi}}(\tG) \cap V| 
    \leq 2|\MM^{\tilde{\pi}}(\tG) \cap E| + |\MM^{\tilde{\pi}}(\tG) \cap \tilde F|
\le 2 \VCopt(G) + |\MM^{\tilde{\pi}}(\tG) \cap \tilde F|.$$
Consider an arbitrary ranking $\tilde\pi$ of $\tilde E$.
Observe that for each $v\in V$, the matching
$\MM^{\tilde{\pi}}(\tG)$ either includes a parallel edge between
$v$ and $v'$ or
it includes a self-loop incident to $v'$.
For every $v' \in V'$, if the lowest rank of self-loops incident to $v'$
is lower than the lowest rank of edges $(v,v')$, then
$\MM^{\tilde{\pi}}(\tG)$ contains one of the self-loops, and does not contain
any parallel edge $(v,v')$. If the ranking $\tilde \pi$ is selected uniformly
at random, the above inequality on ranks does not hold for each vertex independently
 with probability at most $\eps d / 8 d = \eps/8$. Therefore, the expected number of edges
in $\MM^{\tilde{\pi}}(\tG) \cap \tilde F$ is upper bounded by
$\eps n/8$. Without loss of generality, we can assume that $\eps n > 72$, since otherwise
we can read the entire input with only $O(1/\eps^2)$ queries and compute a maximal matching in it. 
It follows from the Chernoff bound that with probability $1-1/20$, 
$|\MM^{\tilde{\pi}}(\tG) \cap \tilde F| \le \eps n/4$.

Observe that given query access to $G$, we can provide query access to $\tG$
(in particular, the edges in $E'$ that are incident to each $v\in V$ can be indexed starting
from $\deg(v)+1$). Every query to $\tG$ can be answered in $O(1)$ time, using $O(1)$
queries to $G$. Therefore, we can simulate an execution of the vertex-cover and the maximal-matching oracles on $\tG$.

Note that the expected number of recursive calls to the maximal matching oracle is bounded for a random vertex $v \in \tilde V$ by $O(d/\eps)$, because the maximum degree and the minimum degree are within a factor of $O(1/\eps)$. 
Also note that since $|V| = |\tilde V|/2$, this expectation for a random vertex $v \in V$ is 
at most twice as much, i.e., it is still $O(d/\eps)$.

For any ranking $\tilde \pi$ of edges in $\tE$,
if we sample $O(1/\eps^2)$ vertices from $V$ with replacement, then the fraction of those in 
$\C^{\tilde{\pi}}(\tG) \cap V$ is within an additive $\eps / 8$ of $|\C^{\tilde{\pi}}(\tG) \cap V
| / |V|$ with probability at least $1 - 1/20$. Let $\mu$ be this fraction of vertices. 
Therefore, we have that
$$\VCopt(G) - \eps n/4\; \le\; \mu \cdot n \;\le\; 2 \VCopt(G) + \eps n/2$$
with probability at least $1-1/10$. 
Thus $(\mu + \eps/4) \cdot n$ is the desired $(2,\eps n)$-estimate.
The expected number of calls to the vertex cover and maximal matching oracles 
is bounded by $O(d/\eps^3)$.  Note that without loss of generality, $\eps \ge 1/4n$,
because any additive approximation to within an additive factor smaller than $1/4$
yields in fact the exact value. Therefore the expected number of calls to the oracles
is bounded by $O(n^4)$, which can be represented with a constant number of machine words 
in the standard binary representation, using the usual assumption that we can address all vertices of the input graph. By applying now Theorem~\ref{thm:running_time},
we obtain an implementation of the algorithm. 
It runs in $O(d/\eps^3 \cdot \log^3(d/\eps))$ time and makes $O(d/\eps^3 \cdot \log^2(d/\eps))$ queries. Moreover, the probability that the implementation diverges from the ideal algorithm is bounded by $1/5$. Therefore, the implementation outputs a $(2,\eps n)$-estimate with probability 
$1 - 1/10 - 1/5 \ge 2/3$.

\begin{corollary}
There is an algorithm 
that 
makes $O(\frac{d}{\eps^3} \cdot \log^3\frac{d}{\eps})$ neighbor and degree
queries, runs in $O(\frac{d}{\eps^3} \cdot \log^3\frac{d}{\eps})$ time, 
and with probability $2/3$, outputs a  $(2,\eps n)$-estimate to the minimum vertex cover size.
\end{corollary}

\subsection{Bounded Average Degree}\label{avg-deg.subsec}

In this section, we assume an upper bound $\bd$ on the average graph degree and 
show an efficient algorithm in this case.\footnote{As shown in~\cite{PR}, 
we don't actually need to know $\bd$ for this
purpose, but it suffices to get a bound that is not much higher than $(4/\eps)\bd$,
and such that the number of vertices with a larger degree is $O(\eps n)$, where such a bound
can be obtained efficiently.} 
To do this, we will transform the graph into a new graph for which
the ratio of the maximum degree to the minimum degree is small.

 Our first transformation is to automatically add high degree vertices
to the cover, and continue by finding a cover for the graph that is induced
by the remaining vertices.
Given a graph $G=(V,E)$ with average degree $\bd$,
let $L$ denote the subset of vertices in $G$ whose degree is greater than $8\bd/\eps$.
Hence, $|L| \le \eps n/8$.
Let $E(L)$ denote the subset of edges in $G$ that
are incident to vertices in $L$, and let $\bG = (\bV,\bE)$ be defined by
$\bV = V\setminus L$ and $\bE = E\setminus E(L)$, so that the maximum degree in 
$\bG$ is at most $8\bd/\eps$.
For any maximal matching $\bM$  in $\bG$ we have that
$$
\VCopt(G) \;\leq\; 2|\bM| + |L| \;\leq\; 2\VCopt(G) + \frac{\eps}{8}n\;.
$$
Thus, the first modification we make to the oracles
is that if the vertex-cover oracle is 
called on a vertex $v$ such that the degree of $v$ is greater
than $(4/\eps)\bd$, then it immediately returns {\sc true}.

 The remaining problem is that when we remove the high degree vertices,
there are still edges incident to vertices with degree at most $(4/\eps)\bd$
whose other endpoint is a high degree vertex, and this is not known until
the appropriate neighbor query is performed.
We deal with this by adding shadow vertices to replace the removed high degree vertices.
At the same time, we increase the minimum
 degree as in the previous subsection.
We now create a graph $\tG = (\bV \cup \tV,\bE \cup \tE)$ as follows.
For every $v \in \bV$, we add to $\tV$ a vertex $v'$ and vertices $v''_i$, where $1 \le i \le \lceil\deg_G(v)/\bd\rceil$ and $\deg_G(v)$ is the degree of $v$ in $G$. Each of these new vertices has $32\bd/\eps$ parallel self-loops.
Moreover, we add $\bd$ parallel edges between $v$ and $v'$. Finally, partition the edges incident
to $v$ in $G$ into $\lceil\deg_G(v)/\bd\rceil$ groups, each of size at most $\bd$. The first group corresponds 
to the first $\bd$ edges on the neighborhood list of $v$, the second group corresponds to the next $\bd$ edges, and so on. Let $E_{v,i} \subset E$ be the set of edges in the $i$-th group.
For every $i$ of interest, we add $|E_{v,i} \cap E(L)|$ parallel edges between $v$ and $v''_i$
and $|E_{v,i} \setminus E(L)|$ parallel self-loops incident to $v''_i$. We add these edges so that we are later able to simulate every query to $\tG$ using a constant number of queries to $G$.

Let us bound the total number of vertices in $\tV$. The number of vertices $v'$ is $|\bV|$.
The number of vertices $v''_{i}$ is bounded by 
$$\sum_{v \in \bV} \left\lceil \frac{\deg_G(v)}{\bd} \right\rceil
\le 
\sum_{v \in \bV} \left(\frac{\deg_G(v)}{\bd} + 1\right)
\le 
\frac{\bd \cdot |\bV|}{\bd} + |\bV| = 2 |\bV|,
$$ 
because $\tV$ has been created by removing vertices with highest degrees in $G$,
and the average degree of vertices in $\tV$ in $G$ cannot be greater than $\bd$,
the initial average degree. This shows that $|\tV| \le 3|\bV|$.

We now repeat an argument from the previous section that despite the additional edges and vertices,
$|\C^{\tilde{\pi}}(\tG) \cap \tV|$ is likely to be a good approximation to $\VCopt(\bG)$
for a random ranking $\tilde{\pi}$. First, $\C^{\tilde{\pi}}(\tG) \cap \tV$ is still a vertex cover for $\bG$, 
so $\VCopt(\bG) \le |\C^{\tilde{\pi}}(\tG) \cap \tV|$. Let $\tilde F$ be the set of edges connecting all $v$ with the corresponding $v'$ and $v''_i$. 
We have 
$$|\C^{\tilde{\pi}}(\tG) \cap \bV| 
    \leq 2|\MM^{\tilde{\pi}}(\tG) \cap \bE| + |\MM^{\tilde{\pi}}(\tG) \cap \tilde F|
\le 2 \VCopt(\bG) + |\MM^{\tilde{\pi}}(\tG) \cap \tilde F|.$$
Observe that if for some of the vertices in $\tV$, the lowest rank of self-loops is lower than the lowest rank of the parallel edges connecting this vertex to the corresponding vertex in $\bV$, then 
one of the self-loops is selected for the maximal matching as opposed to the parallel edges.
The inequality on ranks does not hold with probability at most $\bd/(32\bd/\eps) = \eps/32$ independently for each vertex in $\tV$. It therefore follows from the Chernoff bound that the number of edges in
$\MM^{\tilde{\pi}}(\tG) \cap \tilde F$ is not bounded by $\eps |\tV|/16$ with probability at most
$\exp(-\eps|\tV|/32)$, which is less than $1/20$ if $|\tV| > 100/\eps$, and we can assume that this is the case. (To circumvent the case of $|\tV| \le 100/\eps$, we can modify the algorithm as follows.
If a sampled vertex belongs to a connected component in $\bV$ of size at most $100/\eps$,
then we can read its connected component in $\tG$ and deterministically find a maximal matching that uses only edges in $\bE$ and self-loops in $\tE$. This all takes at most $O(\bd/\eps^2)$ time, which as we see later, we are allowed to spend per each sampled vertex.)
Therefore, we have
$$|\C^{\tilde{\pi}}(\tG) \cap \bV| 
    \le 2 \VCopt(\bG) + \eps|\bV|/4,$$
with probability at least $1 - 1 / 20$.

Observe that given query access to $G$, we can efficiently provide query access to $\tG$.
Degrees of vertices in $\bV$ are the same as in $G$. For associated vertices
in $\tV$ it is easy to compute their degree in $O(1)$ time, using the degree
of the corresponding vertex in $\bV$. To answer neighbor queries for vertices $v$ in $\bV$,
except for the fixed connections to $v'$, it suffices to notice that if the corresponding edge in $G$ is connected to a vertex in $L$, this connection is replaced by a connection to an appropriate vertex $v''_{i}$. Otherwise, the edge remains in $\bE$. For vertices $v''_{i}$ some number of connections can either be a connection to the corresponding $v$ or a self-loop. This can be checked in $O(1)$ time with a single query to the neighborhood list of $v$. All the other edges are fixed. Therefore, we can simulate an execution of the vertex-cover and the maximal-matching oracles on $\tG$.
Answering every query to $\tG$ requires $O(1)$ time and $O(1)$ queries to $G$.
Sampling vertices uniformly at random from in $\bV \cup \tV$ is more involved, but in our algorithm, we only need to sample vertices from $V$, which we assume we can do.

The expected number of recursive calls to the maximal matching oracle is bounded by $O(\bd/\eps^2)$ for a random vertex $v \in \bV \cup \tV$, because the maximum degree and the minimum degree are within a factor of $O(1/\eps)$ and the maximum degree is bounded by $O(\bd/\eps)$. Note that since $3|\bV| \ge |\tV|$, this expectation for a random 
vertex $v \in V$ is at most twice as much, i.e., it is still $O(\bd/\eps^2)$.

For any ranking $\tilde \pi$ of edges in $\tG$,
if we sample $O(1/\eps^2)$ vertices from $V$ with replacement, then the fraction of those for which the oracle answers {\sc true} is within an additive error $\eps / 8$ of 
the total fraction of vertices for which the oracle answers {\sc true}.
with probability $1 - 1/20$. Let $\mu$ be the fraction of sampled vertices. We have 
$$\VCopt(G) - \eps n/8 \le \mu \cdot n \le 2 \VCopt(G) + \eps n/4 + \eps n/8 + \eps n/8$$
with probability $1-1/10$.
Then $(\mu + \eps/8) n$ is the desired $(2,\eps n)$-estimate.
The expected number of calls to the vertex cover and maximal matching oracles is bounded by $O(d/\eps^4)$. As before, without loss of generality, this quantity can be bounded by $O(n^5)$,
which fits into a constant number of machine words. By applying now 
Theorem~\ref{thm:running_time},
we obtain an implementation of the algorithm. 
It runs in $O(d/\eps^3 \cdot \log^3(d/\eps))$ time and makes $O(d/\eps^3 \cdot \log^2(d/\eps))$ queries. Moreover, the probability that the implementation diverges from the ideal algorithm is bounded by $1/5$. Therefore, the implementation outputs a $(2,\eps n)$-estimate with probability 
at least $1-1/5 - 1/10 \ge 2/3$.

\begin{corollary}
There is an algorithm 
that makes $O(\frac{\bd}{\eps^4} \cdot \log^2\frac{\bd}{\eps})$ 
neighbor and degree queries, 
runs in $O(\frac{\bd}{\eps^4} \cdot \log^3\frac{\bd}{\eps})$ time, 
and with probability $2/3$, outputs a  $(2,\eps n)$-estimate to the minimum vertex cover size.
\end{corollary}

} 

\def\Dense{
\subsection{Adapting the Algorithm to the Vertex-Pair Query Model}\label{dense.sec}

The focus of this paper was on designing a sublinear-time algorithm whose access to
the graph  is via degree queries and neighbor queries. In other words, we assumed
that the graph was represented by adjacency lists (of known lengths).
When a graph is dense (i.e., when the number of edges is $\Theta(n^2)$),
then a natural alternative representation is by an adjacency matrix.
This representation supports queries of the form: ``Is there an edge between
vertex $u$ and vertex $v$?'', which we refer to as {\em vertex-pair\/} queries.

We next show how to adapt the algorithm described in the previous section 
to an algorithm that performs vertex-pair queries. 
The query complexity and running time of the algorithm are (with high constant
probability) $\tilde{O}(n/\eps^4)$, which is linear in the average degree for
dense graphs. As in the previous section, the algorithm outputs
(with high constant probability)
a $(2,\eps)$-estimate of the size of the minimum vertex cover.
We recall that the linear lower bound in the average degree~\cite{PR} 
also holds for the case that the average degree is $\Theta(n)$ and 
when vertex-pair queries are allowed.

Given a graph $G = (V,E)$, let $\tG = (\tV,\tE)$ be a supergraph of $G$ that is
defined as follows. For every vertex $v \in V$ whose degree in $G$ is less than\footnote{
If there are no self-loops in the original graph, then the bound is $n-1$.} $n$,
there exists a vertex $v' \in \tV$, where there are $n- \deg_G(v)$ parallel
edges between $v$ and $v'$, and there are $(8/\eps)n$ self-loops incident to $v'$.
As shown in Subsection~\ref{bound-deg.subsec}, with high probability over the choice
of a ranking $\tilde{\pi}$ over $\tG$, we have that 
$|\C^{\tilde{\pi}}(\tG) \cap V| \leq 2 \VCopt(G) + (\eps/4)n$. 

Note that we can emulate neighbor queries to $\tG$ given access to vertex-pair
queries in $G$ as follows. Let the vertices in $G$ be $\{1,\ldots,n\}$. When 
the $j^{\rm th}$ neighbor of vertex $i$ is queried, then the answer to the
query is $j$ when $(i,j) \in E$, and it is $i'$ (the new auxiliary vertex adjacent
to $i$) when $(i,j) \notin E$. The degree of every vertex in $V$ is $n$, so there
is no need to perform degree queries.
Since the ratio between the maximum degree and the minimum degree in
$\tG$ is at most $1/\eps$, and the maximum and average degrees are $O(n/\eps)$,
we obtain an algorithm whose complexity is $\tilde{O}(n/\eps^4)$, as claimed.
}

\ifnum\conf=0
\NearOpt
\Dense
\fi